\documentclass[12pt]{amsart}
\usepackage{amssymb,amsmath}
\usepackage{latexsym}
\usepackage{graphicx}
\usepackage{geometry}                
\usepackage{color}
\usepackage{xcolor}
\usepackage{pdfsync}
\usepackage{fancyhdr}
\usepackage[applemac]{inputenc}
\usepackage[breaklinks,colorlinks,backref]{hyperref}
\hypersetup{
    colorlinks, 
    linktoc=all, 
    linkcolor=red,  
  citecolor=hyptxt,
  urlcolor=blue}
  \hypersetup{
  citebordercolor=,
  filebordercolor=green,
  linkbordercolor=blue
}
\definecolor{hyptxt}{rgb}{0.7, 0.4, 0.9}
\usepackage{bibtopic}

\newcommand{\be}{\begin{equation}}
\newcommand{\en}{\end{equation}}

\newcommand{\bea}{\begin{eqnarray}}
\newcommand{\ena}{\end{eqnarray}}

\newcommand{\ket}[1]{|\kern.3ex#1\kern.3ex\rangle}
\newcommand{\bra}[1]{\langle\kern.3ex #1 \kern.3ex|}
\newcommand{\scalar}[2]{\langle\kern.3ex #1 \kern.3ex|\kern.3ex#2\kern.3ex\rangle}

\newcommand{\C}{\mathbb C}

\newcommand{\NN}{\mathbb N}

\newcommand{\R}{\mathbb R}

\newcommand{\Z}{\mathbb Z}

\newcommand{\SN}{{\mathbb S}}

\newcommand{\h}{{\mathfrak H}}
\newcommand{\hs}{Hilbert space}
\newcommand{\ud}{\,\mathrm{d}}
\newcommand{\dis}{\displaystyle}
\newcommand{\ad}{^{\mbox{\scriptsize $\dag$}}} 
 \newcommand{\ov}{\overline}
\newcommand{\nn}{\nonumber}

\newtheorem{defi}{Definition}[section]

\newtheorem{prop}[defi]{Proposition}

\def\lg{\langle }
\def\rg{\rangle }
\def\llg{\left\langle }
\def\rrg{\right\rangle }
\def\deq{\stackrel{\mathrm{def}}{=}}

\def\adg{a^{\dag}}

\begin{document}
\title[Integral quantizations]{Integral quantizations with two basic examples}
\author{Herv\'e Bergeron and  Jean Pierre Gazeau }
\address{ISMO, UMR 8214 CNRS, Univ Paris-Sud,  France
} \email{herve.bergeron@u-psud.fr}

\address{APC, Univ Paris  Diderot, Sorbonne Paris Cit\'e 75205 Paris, France}\email{gazeau@apc.univ-paris7.fr
}
\address{
CBPF, 22290-180  Rio de Janeiro / RJ
Brasil}\email{gazeaujp@cbpf.fr
}
\date{\today}
\begin{abstract}
The paper concerns integral quantization, a procedure based on operator-valued measure and resolution of the identity.  We insist on covariance properties in the important case where  group representation theory is involved. We also insist on the inherent  probabilistic aspects of this classical-quantum map.  The approach includes and generalizes coherent state quantization.  Two applications based on group representation are carried out. The first one concerns the Weyl-Heisenberg group and the euclidean plane viewed as the corresponding phase space. We show that  a world of quantizations exist, which yield the canonical commutation rule and the  usual quantum spectrum of the harmonic oscillator. The second one concerns the affine group of the real line and gives rise to  an interesting regularization of the dilation origin in the half-plane viewed as the corresponding phase space. 
\end{abstract}

\maketitle
\tableofcontents

\section{Introduction}
\label{sec:intro}

We present in this paper  an  approach to quantization based on operator-valued measures,  comprehended under the generic name of \emph{integral quantization}.  We particularly insist on the probabilistic aspects appearing at each stage of our procedure.
The so-called Berezin or Klauder or yet Toeplitz quantization, and more generally coherent state quantization  are particular (and mostly manageable) cases of this approach. 

The integral quantizations include of course  the ones based on  the Weyl-Heisenberg group (WH), like  Weyl-Wigner and (standard) coherent states quantizations. It is well established that the WH group underlies the canonical commutation rule, a paradigm of quantum physics. Actually, we show that there is a world of quantizations that follow this rule. In addition,  we enlarge the set of objects to be quantized in order to include singular functions or distributions. 

Our approach also includes a less familiar quantization based on the affine group of the real line. This example is illuminating and quite promising in view of applications in various domains of physics where it is necessary to take into account an impenetrable barrier. 

In Section \ref{sec:whatquant} we give a short overview of what we mean by quantization after recalling the basic method in use in Physics. 
The definition of integral quantization is proposed in Section \ref{sec:intquant}. Key examples issued from group representation theory give rise to what we name \emph{covariant integral quantizations}. We then apply the scheme to the Weyl-Heisenberg group in  Section \ref{sec:weylheis}. As stressed on in the above, the freedom allowed by our approach gives rise to  a wide range of quantum descriptions of the euclidean  plane viewed as a phase space, all equivalent in the sense that they yield the canonical commutation rule. 
Moreover, as explained in Section \ref{sec:fctdist}, our approach offers the possibility of dealing with singular functions and distributions and providing in a simple way their respective quantum counterparts. 
Besides the euclidean plane, the half plane can be also viewed as a phase space, and Section \ref{sec:affine} is devoted to the construction(s) of its quantum version through its  unitary irreducible representation intensively used in wavelet analysis. 
In Section \ref{sec:conclu} we conclude our presentation with an agenda of future developments envisioned within the framework presented in this paper. In Appendix \ref{glossDz} is given a list of useful formulae concerning the Weyl-Heisenberg machinery. In Appendix \ref{affqgr} we give a compendium of previous 
works by Klauder where the affine group and related coherent states are also used for quantization of the half-plane, although with a different approach. 
\section{What is quantization?}
\label{sec:whatquant}

In digital signal processing, one generally understands by \textit{quantization} the process of mapping a large set of input values to a smaller set – such as rounding values to some unit of precision. A device or algorithmic function that performs quantization is called a quantizer. In physics or mathematics, the term has a different, and perplexing,  meaning. For instance, one can find in \textit{Wikipedia}:

\begin{quote}
\textit{Quantization is the process of explaining a classical understanding of physical phenomena in terms of a newer understanding known as ``quantum mechanics''. It is also a procedure for constructing a quantum field theory starting from a classical field theory. }
\end{quote}
or  in \cite{kiukas}, 
\begin{quote}
\emph{Quantization can be any procedure that associates a quantum mechanical observable to a given classical dynamical variable.} 
\end{quote}

The standard (canonical) construction is well known. It is based on the replacement, in the expression of classical observables $f(q,p)$, of the conjugate variables $(q,p)$ (i.e., obeying $ \{q,p\} = 1$) by their respective self-adjoint operator counterparts $(Q,P)$. The latter  obey the canonical commutation rule $[Q,P] = i\hbar I$. The substitution has to be followed by a symmetrization. This last step is avoided by the integral version of this method, namely the Weyl-Wigner or ``phase-space'' quantization \cite{weyl28,grossmann76,daub11,daub12,daub13} (see also  \cite{zachos}
and references therein).  The canonical procedure is universally accepted in view of its numerous experimental validations, one of the most famous and simplest one going back to the early period of Quantum Mechanics with the quantitative prediction of the isotopic effect in vibrational spectra of diatomic molecules \cite{herzberg1989}. 
 These data validated the canonical quantization, contrary to the  Bohr-Sommerfeld Ansatz (which predicts no isotopic effect). Nevertheless this does not prove that another method of quantization fails to yield the same prediction (see for instance  \cite{bergayou13}).
  Moreover, canonical or Weyl quantization is difficult, if not impossible, to implement as soon as barriers or other impassable boundaries are present in the model. Similar complications arise when one deals with singular functions $f(q,p)$, even though they are as elementary and familiar as the angle or phase function $\arctan p/q$, an enigmatic question which has been giving rise to a  long debate since the advent of Quantum Mechanics \cite{carnie68,royer96}.  

To be more mathematically precise and still remaining at an elementary or minimal level, quantization is a linear map 
\begin{equation}
\label{quantizmap}
\mathfrak{Q}: \mathcal{C}(X) \mapsto \mathcal{A}(\h)
\end{equation}
 from a vector space  $\mathcal{C}(X) $ of complex-valued functions $f(x)$ on a set $X$ to a vector space  $\mathcal{A}(\h)$ of linear operators $\mathfrak{Q}(f) \equiv A_f$  in some complex Hilbert space $\h$, such that
\begin{itemize}
 \item[(i)] to the function $f=1$ there corresponds the identity operator $I$ on $\h$,
\item[(ii)]  to a real function $f\in \mathcal{C}(X)$ there corresponds a(n) (essentially) self-adjoint operator $A_f$ in  $\h$.
\end{itemize}
Physics puts into the game further requirements, depending on various mathematical structures allocated to $X$ and $\mathcal{C}(X)$,  such as a measure, a topology, a manifold, a closure under algebraic operations, considerations about dynamic evolution..., together with an interpretation in terms of measurements of the elements of $\mathcal{C}(X)$ (resp. $\mathcal{A}(\h)$). These objects are given the status of classical (resp. quantum) physical quantities or \textit{observables}.  In particular, the spectral properties of elements of $\mathcal{A}(\h)$ are at the heart of the quantum measurement protocols. Moreover, one usually requires a non ambiguous classical limit of the quantum physical quantities, the limit operation being associated to a change of scale. It is precisely at this point that both the definitions of quantization, namely Signal Analysis and in Physics,  might be related. 

A warning is in order at this point. According to the standard (von Neumann's) interpretation, \emph{every} self-adjoint operator represents an observable, but this is untenable. On the mathematical side, observables are most often represented by \emph{unbounded} operators  and there will be in general  no dense domain common to all of them. On the physical side, a self-adjoint operator may be interpreted as an observable for one system, but not for another one. 
Therefore one can characterize
a quantum system, with \hs\ $\h$, by a family of \emph{labeled observables} in the sense of Roberts \cite{roberts}, with are
given  both  a physical interpretation (how does one measure it?) and a mathematical definition (as
a self-adjoint operator in $\h$). In addition, these are required to have a common dense invariant domain. 
As a matter of fact, this  is at the basis of the Rigged Hilbert Space formulation of quantum mechanics, developed  in \cite{ant-rhs1,-jpa_rhs,bohmlect}.

\section{Integral quantization}
\label{sec:intquant}
\subsection{General setting}
\label{subsec:set}
The above conditions (i) and (ii) may be easily  fulfilled if one uses the resources offered by the pair \textit{(measure, integration)}. Let $\h$ be a complex Hilbert space and $X\ni x \mapsto {\sf M}(x) \in \mathcal{L}(\h)$ an $X$-labelled  family of bounded operators on $\h$  resolving the identity operator $I$ on $\h$:
 \begin{equation}
\label{Psolveun}
\int_{X}\, {\sf M}(x)\, \ud\nu(x) = I\, , 
\end{equation}
provided that the  equality be valid in a weak sense, which implies $\nu$-integrability for the family ${\sf M}(x)$. (Actually, with the painless requirement on $X$ that it be locally compact, linearity and continuity, w.r.t. some topology,  of the map (\ref{quantizmap}), the latter must be integral quantization where ${\sf M}(x)\, \ud\nu(x)$ is to be replaced by the more general operator-valued measure $\ud \widetilde{{\sf M}} (x)$). If the operators ${\sf M}(x)$ are positive and with unit trace, they are preferably denoted by ${\sf M}(x)= \rho(x)$ in order to comply with the usual notation for a density matrix in quantum mechanics. In this case, and given a positive unit trace operator $\rho_0$, the equality \eqref{Psolveun} allows one to equip the set $X$ with the probability distribution $x\mapsto \mathrm{tr}(\rho_0\,\rho(x))\equiv w(x)$.

The corresponding quantization of complex-valued functions $f(x)$ on $X$ is then defined by the linear map:
 \begin{equation}
\label{formquant}
f\mapsto A_f = \int_{X}\, {\sf M}(x) \, f(x)\, \ud\nu(x)\, . 
\end{equation}
Therefore if ${\sf M}(x)= \rho(x)$ and given a positive unit trace operator $\rho_0$, we obtain the exact classical like formula
\begin{equation}
\label{semiclass}
\mathrm{tr}(\rho_0\,A_f) = \int_X f(x)\, w(x) \, \ud\nu(x)\,. 
\end{equation}
which is to be viewed as an averaging of the original $f$. 

The operator-valued integral \eqref{formquant}  is  understood  as the sesquilinear form, 
\begin{equation}
\h \ni \psi_1, \psi_2 \mapsto B_f(\psi_1,\psi_2)= \int_{X}\, \lg \psi_1|{\sf M}(x)|\psi_2\rg \, f(x)\, \ud\nu(x)\,.
\end{equation}
The form $B_f$ is  assumed to be  defined on a dense subspace of $\h$.  If $f$ is real and at least semi-bounded, and if ${\sf M}$ is positive,  the Friedrich's extension \cite[Thm. X.23]{reedsimon2} of $B_f$ univocally defines a self-adjoint operator. However, if $f$ is not semi-bounded, there is no natural choice of a self-adjoint operator associated with $B_f$. In this case, we can consider directly the symmetric operator $A_f$ enabling us to obtain a self-adjoint extension (unique for particular operators). It becomes necessary to be more explicit about the Hilbert space $\h$ and operator domain(s). The question of what is the class of operators that may be so represented is a subtle one, and its study is far beyond the scope of this paper.

If ${\sf M}(x)= \rho(x)$ and if we dispose of another  $X$-labelled family of positive unit trace operators $X\ni x \mapsto \widetilde{\rho}(x) \in \mathcal{L}^+(\h)$ (it could happen to be the same, or a constant $\rho_0$ like in \eqref{semiclass}) , we can in return build  the classical element in $\mathcal{C}(X) $ 
 \begin{equation}
\label{pbformquant}
A_f \mapsto \check{f}(x) := \int_{X}\, \mathrm{tr}(\widetilde{\rho}(x)\rho(x')) \, f(x')\, \ud\nu(x')\, , 
\end{equation}
provided that the integral be defined. The map $f\mapsto \check{f}$ is  a generalization of the Segal-Bargmann transform (\cite{stenzel}). Furthermore, the function or \textit{lower symbol}\footnote{A name introduced by Lieb, by opposition to  \textit{upper symbol} for designating the original $f$. Berezin used the names  \textit{covariant symbol} and  \textit{contravariant symbol} respectively} $\check{f}$ may be viewed as a semi-classical representation of the operator $A_f$.  
It is at this point that the mentioned classical limit gets a concrete meaning, given a scale parameter $\epsilon$ and a distance $d(f,\check{f})$:
\begin{equation}
\label{classlim1}
d(f,\check{f})\to 0 \quad \mbox{as}\quad \epsilon \to 0\,. 
\end{equation}
One of the interesting aspects  of the integral quantization scheme is precisely to analyze the quantization issues, particularly the spectral properties of operators $A_f$, from functional properties of its lower symbol  $\check{f}$.

Another appealing aspect of the above approach lies in its capability to quantize constraints. Suppose that the measure set $(X,\nu)$ is also a smooth manifold of dimension $n$, on which is defined the space $\mathcal{D}'(X)$ of distributions  as the topological dual of the (LF)-space $\Omega_c^n(X)$ of compactly supported $n$-forms on $X$ \cite{grosser08}. 
Some of these distributions, e.g.  $\delta(u(x))$, express geometrical constraints. Extending the map \eqref{formquant} yields the quantum version $A_{\delta(u(x))}$ of these constraints. 

A different starting point for quantizing constraints, more in Dirac's spirit \cite{dirac64} 
and intensively employed in (Loop) Quantum Gravity and Quantum Cosmology (see \cite{bom-klau,
dewitt,kamin,rovelli-lect,wieland12} 
and references therein)
would consist in quantizing the function $u \mapsto A_u$ and determining the kernel of the operator  $A_u$. Both methods are obviously not equivalent, except for a few cases. This question of equivalence/difference gives rise to controversial opinions in fields like quantum gravity or quantum cosmology. Elementary examples illustrating this difference are worked out  in \cite{balfrega13}.

\subsection{Covariant integral   quantization}
\label{subsec:covq}
Lie group representation theory offers a wide range of possibilities for the building of explicit integral quantizations. Let $G$ be a Lie group with left Haar measure $\ud\mu(g)$, and let $g\mapsto U(g)$ be a unitary irreducible representation of $G$ in a Hilbert space $\h$. Let ${\sf M}$ a bounded operator on $\h$. Suppose that the  operator 
\begin{equation}
\label{intgrR}
R:= \int_G  \, {\sf M}(g)\,\ud\mu(g) \, , \quad  {\sf M}(g):= U(g)\, {\sf M}\, U^{*}(g)\, , 
\end{equation}
is defined in a weak sense. From the left invariance of $\ud\mu(g)$ we have  $U(g_0)\,R\, U^{*}(g_0)= \int_G \ud\mu(g) \, {\sf M}(g_0g) = R$ and so $R$ commutes with all operators $U(g)$, $g\in G$.  Thus Schur's Lemma \cite[Chap.5, \S 3]{-barracz} 
 guarantees that $R$ is a multiple of the identity, $R= c_{ {\sf M}}I$. The constant  $c_{ {\sf M}}$ can be given by
\begin{equation}
\label{calcrho}
c_{ {\sf M}} = \int_G  \, \mathrm{tr}\left(\rho_0\, {\sf  M}(g)\right)\, \ud\mu(g)\, ,
\end{equation}
where the unit trace positive operator $\rho_0$ is suitably chosen in order to make the integral convergent. 
Integrating this constant into the measure, we obtain the resolution of the identity obeyed by the family of operators $U(g)\, {\sf M}\, U^{\dag}(g)$:
\begin{equation}
\label{Resunityrho}
\int_G \, {\sf M}(g) \,\ud \nu(g) = I\,, \quad \ud \nu(g):= \ud\mu(g)/c_{ {\sf M}}\, . 
\end{equation} 
For instance, in the case of a square-integrable UIR $U$, let us pick a unit vector $|\eta\rg$ for which $c_{ {\sf M}} \equiv c(\eta)= \int_G \ud\mu(g) \, \vert \lg \eta| U(g)\eta\rg\vert^2$$|\eta\rg < \infty$, i.e  is an admissible unit vector for $U$ \cite{aag00}. With  $ {\sf M }:= |\eta\rg\lg\eta|$ we recover the resolution of the identity provided by the family of  states:
\begin{equation}
\label{Gcs}
|\eta_g\rg= U(g)|\eta\rg\, ,
\end{equation}
\begin{equation}
\frac 1{c(\eta )}\int_{G}\vert\eta_{g}\rangle\langle\eta_{g}\vert {\ud} \mu (g) = I\,. 
\label{sqintrep2}
\end{equation} 
Vectors $|\eta_g\rg$ are named (generalized) coherent or wavelet for the group $G$, depending on the context. 

The construction leading to \eqref{Resunityrho} provides an integral  quantization of complex-valued functions on the group 
\begin{equation}
\label{Gintquant}
f\mapsto A_f = \int_G\,  {\sf M(g)}\,f(g)\, \ud \nu(g) \, . 
\end{equation}
Moreover, this quantization is covariant in the sense that
\begin{equation}
\label{covintquant}
U(g) A_f U^{\dag}(g) = A_{U(g)f}\, ,  
\end{equation}
where $(U(g)f)(g'):= f(g^{-1}g')$ is the regular representation if $f\in L^2(G,\ud\mu(g))$.

If $ {\sf M} = \rho$, with the most immediate choice $\widetilde{\rho}= \rho$, we obtain the generalization of the Berezin or heat kernel transform on $G$: 
\begin{equation}
\label{Gberezin}
f\mapsto \check{f}(g) := \int_{G}\, \mathrm{tr}(\rho(g)\,\rho(g')) \, f(g')\,\ud\nu(g')\, . 
\end{equation}

Within the same group theoretical framework, and in the absence of square-integrability of the unitary representation $U$ over the whole group, there exists  a definition  of square-integrability of $U$  and related coherent states  with respect to a left coset manifold $X= G/H$, with $H$ a closed subgroup of $G$,  equipped with a quasi-invariant measure $\nu$. This opens another world of possible covariant quantizations (see \cite{aag00} for notations, details, and references therein). 

For a global Borel section 
$\sigma : X \rightarrow G$ of the group, let $\nu_{\sigma}$ be the unique 
quasi-invariant measure defined by 
\begin{equation}
\label{stanquasinvmeas1}
 {\ud}\nu_{\sigma}(x) = \lambda (\sigma (x), x){\ud}\nu (x) \, ,  
\end{equation}
where $\lambda$ is defined by
\begin{equation}
\label{lambdadef}
\ {\ud}\nu (g^{-1}x)=\lambda (g, x){\ud}\nu (x) ,\;\; (\forall g \in G)\,.
\end{equation}

Let $U$ be a UIR 
which is square integrable
$\mbox{\rm mod}(H)$ and $n$ vectors $\eta^{i}$ which are admissible  $\mbox{\rm mod}(H)$.
With $F = \sum_{i=1}^n \vert\eta^{i} \rangle \langle 
\eta^{i}\vert$, $F_{\sigma}(x) = U(\sigma (x))FU(\sigma (x))^{\dag}$, 
we have the resolution of the identity
\be
\frac{1}{c_F}\int_X \; F_{\sigma}(x) \, {\ud}\nu_{\sigma}(x) = I , \label{resunimodH}
\en
the integral converging weakly. If $\mathrm{tr}F = 1$, the constant $c_F$ is given by the integral 
$c_F= \int_X  \, \mathrm{tr}\left(F\,F_{\sigma}(x)\right)\, {\ud}\nu_{\sigma}(x)$.

Consider now the sections $\sigma_{g}: X \rightarrow G, \;\; g \in G$, which are 
covariant translates of $\sigma$ under $g$:
\be
   \sigma_{g}(x) = g\sigma (g^{-1}x) = \sigma (x)h(g, g^{-1}x)\, .
\label{translsection2}
\en
Here $h$ is the cocycle defined by 
\begin{equation}
g\sigma (x) = \sigma (gx)h(g,x) \quad \mbox{with} \quad
h(g,x)  \in H\, .
\label{cocyclecond2}
\end{equation}
 Let $\nu_{\sigma_g}$
be the  measure  ${\ud}\nu_{\sigma_g}(x) := \lambda (\sigma_{g} (x), x)\,{\ud}\nu$, again
constructed  using (\ref{stanquasinvmeas1}), and $F_{\sigma_g}(x) = 
U(\sigma_{g}(x))FU(\sigma_{g}(x))^{\dag}$. If $U$ is square integrable  $\mathrm{mod}(H, \sigma )$, 
there is a general covariance property enjoyed, when they are properly defined,  by the operators 
\begin{equation}
\label{quantizcovmodH1}
A_f = 
{c_F}^{-1}\int_X \; F_{\sigma}(x) \, f(x)\, {\ud}\nu_{\sigma}(x)
\end{equation}
 which are the quantized versions  of functions $f(x)$ on $X$. 
\begin{equation}
\label{quantizcovmodH2}
 U(g) A_f U(g)^{\ast} = A^{\sigma_g}_{U_l(g)f}\, , \quad A^{\sigma_g}_f:= \frac{1}{c_F}\int_X F_{\sigma_g}(x) f(x) \ud \nu_{\sigma_g}(x)\,. 
\end{equation} 
Of course, it is possible to establish similar results by replacing the operator $F$ by a more general bounded operator ${\sf M}$ provided integrability and weak convergence hold in the above expressions.

\section{Exemple: Weyl-Heisenberg covariant integral quantization(s)}
\label{sec:weylheis}

\subsection{The  group background}

The formalism and quantization procedure introduced above allow one to considerably encompass the Berezin-Klauder-Toeplitz quantization based on standard (i.e. Glauber) coherent states (see for instance \cite{alienglis2005,gazeaubook09} and references therein). The group $G$ is the Weyl-Heisenberg group $\mathrm{G}_{\rm WH}$ which is a central extension of the group of translations of the two-dimensional Euclidean plane. In classical mechanics the latter is viewed as the phase space for the motion of a particle on the real line. The UIR in question is the unitary representation of $\mathrm{G}_{\rm WH}$ which integrates the canonical commutation rule of Quantum Mechanics, $[Q,P]=i\hbar I$. Forgetting about physical dimensions ($\hbar = 1$), an arbitrary  
element $g$ of  $\mathrm{G}_{\rm WH}$ is of the form   
\begin{equation}
\label{weylheispar}
 g = (\theta , q,p)\equiv (\theta,z), \quad \theta \in {\mathbb R}\,, \quad (q,p)  \in  {\mathbb R}^{2}  \ \mbox{and}\ z := \frac{q+ip}{\sqrt{2}} \in \C\, . 
\end{equation}      
with multiplication law,  
\begin{equation}
\begin{aligned}
g_{1}g_{2} &= (\theta_{1} + \theta_{2} + \xi ((q_{1},p_{1}) ; (q_{2},p_{2})), \;  
           q_{1}+q_{2},\; p_{1}+p_{2}) \\
           & = (\theta_{1} + \theta_{2} + \frac{1}{2}\mathrm{Im}\,z_1\bar z_2, \;  
           z_1 + z_2)\, ,  
\label{WHgroupmult} 
\end{aligned}
\end{equation}
where $\xi$ is the multiplier function $\xi ((q_{1},p_{1}) ; (q_{2},p_{2})) = \frac 1{2} (p_{1}q_{2} - p_{2}q_{1}). 
$
Any infinite-dimensional UIR, $U^{\lambda}$, of $\mathrm{G}_{\rm WH}$ is characterized by a real number $\lambda \neq 0$
(in addition, there are also degenerate, one-dimensional, UIRs corresponding to $\lambda = 0$, but they are 
irrelevant here \cite{-perel2}) and   may be realized on the same Hilbert space $\h$, as the one carrying an irreducible  
representation of the CCR:  
\begin{equation}
 U^{\lambda}(\theta , q, p)   =   e^{i\lambda\theta} U^{\lambda}(q,p)  
                             :=  e^{i\lambda(\theta - pq/2)}
                             e^{i\lambda pQ} e^{- i\lambda qP}. 
\label{WHgrouprep1} 
\end{equation}
If $\h = L^{2}({\mathbb R} , {\ud}x)$, these operators are defined by the action 
\begin{equation}
(U^{\lambda}(\theta , q, p)\phi )(x) = e^{i\lambda\theta}
    e^{i\lambda p(x- q/2)}\phi (x-q), \qquad  
    \phi \in L^{2}({\mathbb R} , {\ud}x). 
\label{WHgrouprep2} 
\end{equation}
Thus, the three operators, $I, Q, P$, appear now  
as the infinitesimal generators of this representation and are realized as:
\begin{equation}
 (Q\phi)(x) = x\phi (x), \quad  (P\phi)(x) = 
   -\frac i\lambda \frac {\partial\phi}{\partial x}(x), \qquad [Q,P] = \frac i\lambda I.
\label{genrepWHgroup}
\end{equation}
For our purposes, we take for $\lambda$ the specific value, 
$\lambda =1/\hbar = 1$, and simply write $U$ for the corresponding 
representation.  

\subsection{Weyl-Heisenberg standard and non-standard CS}
Picking $H$ as the phase subgroup  $\mathrm{\mathrm{\Theta}}$
 (the subgroup of elements $g = (\theta , 0,0), \; \theta \in {\mathbb R})$ 
the measure space to be considered here is the left coset space $X\equiv \mathrm{G}_{\rm WH}/\mathrm{\Theta}$ . It is identified with  the euclidean plane
${\mathbb R}^{2}$ and a general element in it is parametrized by $(q,p)$, consistently to \eqref{weylheispar}. In terms of this  
parametrization, $\mathrm{G}_{\rm WH}/\mathrm{\Theta}$ carries the \emph{invariant} measure 
\be 
    {\ud}\nu (q,p) = \frac {{\ud}q{\ud}p}{2\pi} = \frac{\ud^2 z}{\pi}\, . 
\label{WHinvmeas} 
\en 
The function 
\be 
   \sigma : \mathrm{G}_{\rm WH}/\mathrm{\Theta} \rightarrow \mathrm{G}_{\rm WH}, \quad  \sigma (q,p) = (0 , q, p), 
\label{WHsection} 
\en 
then defines a \emph{section} in the group $\mathrm{G}_{\rm WH}$, now viewed as a \emph{fibre bundle}, 
over the base space $\mathrm{G}_{\rm WH}/\mathrm{\Theta}$, having fibres isomorphic to $\mathrm{\Theta}$. Picking a vector $\eta \in \h$,  
a family of  Weyl-Heisenberg  CS (they are called Gabor states in time-frequency signal analysis) is the set,
\be 
   {\mathfrak S}_{\sigma} = \{ \eta_{\sigma (q, p)}^{s} =  
              U(\sigma (q, p))\eta \;\vert\; (q,p) \in \mathrm{G}_{\rm WH}/\mathrm{\Theta} \},  
\label{cancohst2} 
\en 
and the operator integral in (\ref{resunimodH}) reads in this case 
\be 
  \int_{\mathrm{G}_{\rm WH}/\mathrm{\Theta}}\vert\eta_{\sigma (q, p)} \rangle 
         \langle \eta_{\sigma (q, p)}\vert\;{\ud}\nu (q,p) = I . 
\label{resolid2} 
\en 
In other words, the CS $\eta_{\sigma (q, p)}$ are labelled by the  
points $(q,p)$ in the \emph{homogeneous space} $\mathrm{G}_{\rm WH}/\mathrm{\Theta}$ of the Weyl-Heisenberg   group, 
 \index{group (explicit)!Weyl-Heisenberg $\mathrm{G}_{\rm WH}$}
and they are obtained by the action of the unitary operators   
$U(\sigma (q, p))$  of a UIR of $\mathrm{G}_{\rm WH}$, on a fixed vector $\eta \in \h$.  These vectors are standard CS when $\eta$ is the gaussian in position representation. 
The resolution of the identity equation (\ref{resolid2}) is then a statement of the  
\emph{square-integrability} of the UIR,  $U$, with respect to the homogeneous   space $\mathrm{G}_{\rm WH}/\mathrm{\Theta}$. 

In the sequel, we will stick to the notations involving the complex variable $z$. Accordingly, the above operator $U(\sigma (q, p))$ is  denoted $D(z)$ and named \textit{displacement operator}. It is given in terms of the lowering $a:= (Q+iP)/\sqrt{2}$ and raising $\adg := (Q-iP)/\sqrt{2}$ operators, $[a,\adg]= I$,  by
\begin{equation}
\label{displop}
D(z) = e^{{\dis z a\ad - \overline{z} a}} \equiv  U(\sigma (q, p))\, .
\end{equation}
On a more abstract level,  lowering and raising operators act in some separable Hilbert space $\h$ with orthonormal basis $\{|e_n\rg\}$, $n\in \NN$: $a|e_n\rg = \sqrt{n}|e_{n-1}\rg$, $a|e_0\rg = 0$, $a^\dag|e_n\rg = \sqrt{n+1}|e_{n+1}\rg$.  A list of basic definitions and of important properties  of $D(z)$, useful for the sequel, are given in Appendix \ref{glossDz}
\subsection{Weyl-Heisenberg integral quantization}
\subsubsection{With a generic weight function}
Let $\varpi(z) $ be a function on the complex plane obeying the condition
\begin{equation}
\label{varpi0}
\varpi(0) = 1\, , 
\end{equation}
and defining a bounded operator ${\sf M}$ on $\h$ through the operator-valued integral
\begin{equation}
\label{opMvarpi}
{\sf M}= \int_{\mathbb{C}} \varpi(z) D(z)\,  \frac{\ud^2 z}{\pi}\, .
\end{equation}
Then, the family ${\sf M}(z):= D(z) {\sf M}D(z)^\dag$ of displaced operators under the unitary action $D(z)$ resolves the identity 
\begin{equation}
\label{residMz}
\int_{\mathbb{C}} \, {\sf M}(z) \,\frac{\ud^2 z}{\pi}= I\, . 
\end{equation}
It is a direct consequence of the  property $D(z) D(z') D(z)^\dag = e^{z\ov{z}' -\ov{z} z'} D(z')$,  of the formula
 \begin{equation}
\label{sympFour1}
\int_{\mathbb{C}} e^{ z \bar \xi -\bar z \xi} \,  \frac{\ud^2 \xi}{\pi} = \pi \delta^{2}(z)\, ,
\end{equation}
and of the condition (\ref{varpi0}) with $D(0)= I$.
The resulting quantization map is given by
\begin{equation}
\label{eqquantvarpi}
f \mapsto A_f = \int_{\mathbb{C}} \, {\sf M}(z) \, \, f(z) \,\frac{\ud^2 z}{\pi}\, . 
\end{equation}
Using the symplectic Fourier transform defined, in agreement with  Eq. (\ref{sympFour1}), by
\begin{equation}
\label{symFourTr}
\hat{f}(z)=\int_{\mathbb{C}} e^{ z \bar \xi -\bar z \xi} f(\xi)\,  \frac{\ud^2 \xi}{\pi}\, , 
\end{equation}
we have the alternative expression
\begin{equation}
\label{quantvarpi1}
A_f = \int_{\mathbb{C}} \varpi(z) \, D(z)\, \hat{f}(-z)\, \frac{\ud^2 z}{\pi} \, .
\end{equation}
A first property 
In accordance with (\ref{sympFour1}) and (\ref{symFourTr}),  the operator $\sf M$ is recovered through the quantization of the Dirac distribution on $\C$:
\begin{equation}
\label{quantdelta}
\sf M = A_{\pi \delta^{(2)}(z)}\, .
\end{equation}
This property extends a result found by Grossmann \cite{-gros_par} within the more restrictive Weyl-Wigner quantization framework and will be comprehensively examined in Section \ref{sec:fctdist}.
 
The covariance property of the above quantization reads as
\begin{equation}
\label{covquant}
A_{f(z-z_0)} = D(z_0) A_{f(z)} D(z_0)^\dag\, .
\end{equation}
Moreover we  have
\begin{equation}
\label{quantvarpi2}
\ A_{f(-z)} = {\sf P} A_{f(z)} {\sf P}\, , \, \forall \,f\  \  \iff \   \varpi(z)=\varpi(-z)\, , \,\forall \,z\, ,
\end{equation}
where ${\sf P} = \sum_{n=0}^{\infty}(-1)^n[e_n\rg\lg e_n|$ is the parity operator defined in \eqref{parity},
\begin{equation}
 \ A_{\overline{f(z)}} = A_{f(z)}^\dag\, , \forall \,f   \ \iff \   \overline{\varpi(-z)}=\varpi(z)\, , \, \forall \,z\, .
\end{equation}
More generally, let us define the unitary representation $\theta \mapsto U_{\mathbb{T}}(\theta)$ of the torus $\SN^1$ on the Hilbert space $\mathcal{H}$ as the diagonal operator  $U_{\mathbb{T}}(\theta)|e_n\rg = e^{i (n + \nu) \theta}|e_n\rg$, where $\nu$ is arbitrary real. We easily infer from the matrix elements  \eqref{matelD} of $D(z)$ the  rotational covariance property
\begin{equation}
\label{rotcovD}
U_{\mathbb{T}}(\theta)D(z)U_{\mathbb{T}}(\theta)^{\dag} = D\left(e^{i\theta}z\right)\, , 
\end{equation}
and its immediate consequence on the nature of $\sf M $ and the covariance of $A_f$,
\begin{equation}
\label{rotcovAf}
U_{\mathbb{T}}(\theta)A_f U_{\mathbb{T}}(-\theta)= A_{T(\theta)f}  \ \iff \  \varpi\left(e^{i\theta}z\right)= \varpi(z) \, , \, \forall \,z\,, \theta  \ \iff \  \sf M \ \mbox{diagonal}\, , 
\end{equation}
where $T(\theta)f(z):= f\left(e^{-i\theta} z\right)$.

\subsubsection{Regular and isometric quantizations}

The  quantization map \eqref{eqquantvarpi} is said to be \emph{regular} (in the sense that  it yields the canonical commutation rule $[a,a^{\dag}] = I$),  if the weight function $\varpi$ verifies the two conditions
\begin{enumerate}
\item[(i)] $\varpi(-z)=\varpi(z)$ (parity),
\vspace*{1mm}\item[(ii)] $\overline{\varpi(z)}=\varpi(z)$ (reality).
\end{enumerate}
In that case we have 
\begin{equation}
\label{regquant1}
A_{z} = a\, , \quad  A_{\overline{f(z)}} = A_{f(z)}^\dag\, . 
\end{equation}
Moreover, the quantization of the canonical variables is equivalent to the canonical quantization:
\begin{equation}
\label{regquantqp1}
A_q = \frac{a + \adg}{\sqrt{2}} := Q \, , \, A_p= \frac{a-\adg}{i \sqrt{2}} := P\,.
\end{equation}
We say that the quantization mapping $f \mapsto A_f$ is \emph{isometric} if $| \varpi(z) |=1$ for all $z$. In this case, and only for it, we have
\begin{equation}
\mathrm{tr} (A_f^\dag A_f) = \int_\C |f(z)|^2 \frac{\ud^2z}{\pi}\, ,
\end{equation}
which means that the mapping $f \mapsto A_f$ is invertible (the inverse is given by a trace formula).

\subsubsection{Elliptic regular quantizations}
It is noticeable that the quantization map (\ref{eqquantvarpi})  includes the normal (as a limit case), anti-normal and Wigner-Weyl (i.e. canonical) quantizations \cite{bergayou13}.
More precisely, 
the quantization map $f \mapsto A_f$ is said to be \emph{elliptic regular}  if the weight function $\varpi$ satisfies, like in \eqref{rotcovAf}, the isotropic condition $\varpi(z) \equiv w(|z|^2)$ with $w: \R \mapsto \R$ (i.e., it leads to  a regular quantization). 
The normal, Wigner-Weyl and anti-normal (i.e., CS) quantizations
correspond to a specific  choice of the weight function $\varpi$. Indeed, inspired by \cite{cahillglauber69}, we choose  
\begin{equation}
\label{standvarpi}
\varpi_s(z) = e^{s |z|^2/2}\, , \quad \mathrm{Re}\; s<1\, . 
\end{equation}
Since this function is isotropic in the complex plane, the resulting operator ${\sf M}\equiv {\sf M}_s$ is diagonal. 
From the expression (\ref{matelD}) of the matrix elements of $D(z)$,  involving associated Laguerre polynomials,
and the integral \cite{magnus66}, 
\begin{equation}
\label{integasslag}
\int_{0}^{\infty} \, e^{-\nu x}\, x^{\lambda}\, L_{n}^{\alpha}(x)\ud x =\frac{\Gamma(\lambda + 1)\Gamma(\alpha+n+1)}{n!\, \Gamma(\alpha + 1)}\nu^{-\lambda -1 }{}_{2}F_1(-n,\lambda + 1; \alpha + 1; \nu^{-1})\, , 
\end{equation}
we get the diagonal elements of  ${\sf M}_s$:
\begin{equation}
\label{diagMs}
\lg e_n|{\sf M}_s|e_n\rg = \frac{2}{1-s}\,\left( \frac{s+1}{s-1} \right)^n\, , 
\end{equation}
and so
\begin{equation}
\label{defMs}
{\sf M}_s= \int_{\mathbb{C}}\;\varpi_s(z) D(z) \,\frac{{\ud}^2z}{\pi }= \frac{2}{1-s} \exp \left\lbrack\left( \log \dfrac{s+1}{s-1}\right) a^\dag a \right\rbrack\,.
\end{equation}
Then $s=-1$ corresponds to the CS  (anti-normal) quantization, since 
\begin{equation*}
{\sf M}_{-1}= \lim_{s\to -1_{-}} \dfrac{2}{1-s} \exp \left\lbrack\left( \ln \dfrac{s+1}{s-1}\right) a^\dag a \right\rbrack = |e_0\rg\lg e_0|\, , 
\end{equation*}
and so 
\begin{equation}
\label{csquants-1}
A_f = \int_{\mathbb{C}} \,  D(z){\sf M}_{-1}D(z)^{\dag} \,f(z) \, \frac{\ud^2 z}{\pi}= \int_{\mathbb{C}} \,  |z\rg\lg z| \, f(z) \,\,\frac{\ud^2 z}{\pi}\, .
\end{equation}
The choice $s=0$ corresponds to the Wigner-Weyl quantization since, from Eq. (\ref{defMs}), 
$
\sf M_0= 2\sf P
$,
and so 
\begin{equation}
\label{wigweylquant}
A_f = \int_{\mathbb{C}} \, D(z) \,2{\sf P}\, D(z)^{\dag}\,  f(z) \,  \,\frac{\ud^2 z}{\pi}\, .
\end{equation}
The case $s=1$ is the normal quantization in an asymptotic sense. 

The parameter $s$ was originally introduced  by Cahill and Glauber in \cite{cahillglauber69,cahillglauber69_2} (see \cite{agarwal-wolf70_1,agarwal-wolf70_2,agarwal-wolf70_3,royer96} for further related developments) where they discuss the problem of expanding an arbitrary operator as an ordered power series in  $a$ and $\adg$.  They associate with every complex number $s$ a unique way of ordering all products of these operators. Normal ordering, antinormal ordering (yielded by CS quantization), and symmetric ordering (yielded by Weyl quantization) correspond to the values $s=+1$, $s=-1$, and $s=0$ respectively. Actually, Cahill and Glauber were not  interested in the question of quantization itself. They start from a symmetric operator-valued series $A(a,\adg)$ in terms of powers of $a$ and $\adg$, without considering their classical counterpart from which they could be built from a given quantization procedure. They ask about their mathematical relations  when a certain $s$-dependent order is chosen. Nevertheless, their work allows to give, to a certain extent, a unified view of different quantizations  of the functions on $\C$.  

Now, an important point favoring in particular the CS quantization is that (from (\ref{diagMs})) the operator ${\sf M}_s$ is positive trace class for $s \leq -1$ (it is just trace class if $\mathrm{Re}\;s<0$). Its trace is equal to 1. 
Therefore, for any real $s\leq -1$, ${\sf M}_s \equiv \rho_s$ is a density operator. Its associated quantization is given a consistent  probabilistic interpretation as is suggested  in \eqref{semiclass} and confirmed from the fact that  the operator-valued measure 
 \begin{equation}
\label{spovs}
\C \supset \Delta \mapsto  \int_{\Delta\in \mathcal{B}(\C)}  D(z){\sf M}_sD(z)^{\dag}\,\dfrac{\ud^2 z}{\pi} \, , 
\end{equation}
is a \underline{positive} operator-valued measure. Moreover,  the form of (\ref{defMs}) suggests that for $s\leq -1$ and given an elementary quantum energy, say $\hbar \omega$, the $s$-dependent  temperature 
\begin{equation}
\label{tempbolt}
\ln \frac{s+1}{s-1} = -\frac{\hbar \omega}{k_BT}\Leftrightarrow s = - \coth\frac{\hbar \omega}{2k_B T}
\end{equation}
 is intrinsically  involved in the quantization procedure of the phase space $\C$ through the Boltzmann-Planck like density operator
 \begin{equation}
\label{plboltrho}
\rho_s= \left( 1- e^{-\tfrac{\hbar \omega }{k_B T}}\right)\sum_{n=0}^{\infty} e^{-\tfrac{n\hbar \omega }{k_B T}}|e_n \rg\lg e_n|\,. 
\end{equation} 
Interestingly, the temperature-dependent operators $\rho_s(z) = D(z) \, \rho_s\, D(z)^{\dag}$ defines a  Weyl-Heisenberg covariant  family of POVM's on the phase space $\C$, the null temperature limit case being the  POVM built from standard CS. Introducing in such a way a temperature suggests that we quantize the classical phase space by taking into account a kind of irreducible noise. 
\subsubsection{Elliptic regular quantizations that are isometric}
An elliptic regular quantization is isometric iff $\varpi(z)=w(|z|^2) \in \{ -1,+1 \}$. A simple example is given by the family $\varpi_\alpha$:
\begin{equation}
\varpi_\alpha(z)=2 \theta(1-\alpha |z|^2)-1
\end{equation}
where $\theta$ is the Heaviside function. The Wigner-Weyl quantization is a special case ($\alpha=0$). 

\subsubsection{Hyperbolic regular quantizations}

We say that the quantization map $f \mapsto A_f$ is   hyperbolic regular if the weight function $\varpi$ verifies $\varpi(z)\equiv
 {\sf m}(\mathrm{Im}\; (z^2))$ with ${\sf m}: \R \mapsto \R$ (i.e. yields a regular quantization). 
In that case we have
\begin{equation}
A_{f(q)}=f(Q)\, , \quad A_{f(p)} = f(P) \,.
\end{equation}
One observes that the Wigner-Weyl (i.e. canonical) quantization (${\sf m}(u)=1$) is not the unique one providing these relations.

\subsubsection{Hyperbolic regular quantizations that are isometric}
An hyperbolic regular quantization is isometric iff  ${\sf m}(u) \in \{ -1,+1 \}$. A simple example is given by the family $\varpi_\alpha$:
\begin{equation}
\varpi_\alpha(z)=2 \theta(1-\alpha \mathrm{Im}\; (z^2))-1\,. 
\end{equation}
The Wigner-Weyl quantization is a special case ($\alpha=0$). 
The above relations imply that all the main properties of the Wigner-Weyl quantization scheme are verified in that case. 

\subsection{Quantum harmonic oscillator according to $\varpi$}
\label{QOintq}
Given a general weight function $\varpi$, the quantization of the classical harmonic oscillator energy $\vert z \vert^2 = (p^2 + q^2)/2$ yields the operator
\begin{equation}
\label{quantosc1}
A_{\vert z \vert^2} = \varpi(0) \adg a + \left.\partial_{z}\varpi\right\vert_{z=0}\,a -  \left.\partial_{\bar z}\varpi\right\vert_{z=0}\,\adg + \frac{\varpi(0)}{2} -  \left.\partial_{z}\partial_{\bar z}\varpi\right\vert_{z=0}\,.
\end{equation}
In the case of a regular quantization, we obtain the operator
\begin{equation}
A_{\vert z \vert^2} = \adg a + \frac{1}{2}  -  \left.\partial_{z}\partial_{\bar z}\varpi\right\vert_{z=0}\,.
\end{equation}
Furthermore,
\begin{eqnarray}
A_{q^2} = Q^2 - \left.\partial_{z}\partial_{\bar z}\varpi\right\vert_{z=0} + \frac{1}{2} \left( \left.\partial_{z}^2 \varpi\right\vert_{z=0}+ \left.\partial_{\bar z}^2 \varpi\right\vert_{z=0}\right) \\
A_{p^2}= P^2 - \left.\partial_{z}\partial_{\bar z}\varpi\right\vert_{z=0} - \frac{1}{2} \left( \left.\partial_{z}^2 \varpi\right\vert_{z=0}+ \left.\partial_{\bar z}^2 \varpi\right\vert_{z=0}\right)
\end{eqnarray}

One observes that the difference between the ground state energy of the quantum harmonic oscillator, namely $E_0= 1/2- \left.\partial_{z}\partial_{\bar z}\varpi\right\vert_{z=0}$, and the minimum of the quantum potential energy, namely $E_m=[\min(A_{q^2}) + \min(A_{p^2})]/2 = - \left.\partial_{z}\partial_{\bar z}\varpi\right\vert_{z=0}$  is $E_0-E_m=1/2$. It is independent of the particular (regular) quantization chosen. 

In the exponential Cahill-Glauber case $\varpi_s(z) = e^{s\vert z \vert^2/2}$ the above operators reduce to 
\begin{equation}
\label{quantosc2}
A_{\vert z \vert^2} = \adg a + \frac{1-s}{2}\,, A_{q^2}=Q^2-\frac{s}{2} \, A_{p^2}=P^2 - \frac{s}{2} \,.
\end{equation}
It has been proven in \cite{bergayou13} that, once restored physical dimensions, these constant shifts in energy are inaccessible to measurement. 

\subsection*{Variations on the Wigner function}

The Wigner function is (up to a constant factor) the Weyl transform of the quantum-mechanical density operator. For a particle in one dimension it takes the form (in units $\hbar=1$)
\begin{equation}
\label{wigtransf1}
\mathfrak{W}(q,p)= \frac{1}{2\pi}\int_{-\infty}^{+\infty}\, \left\lg q-\frac{y}{2}\right\vert\left. \rho| q +\frac{y}{2}\right\rg\, e^{ipy}\, \ud y\,. 
\end{equation}
Adapting this definition to the present context, and  given an operator $A$, the corresponding  Wigner function is defined as
\begin{equation}
\label{wigA}
\mathfrak{W}_A(z) = \mathrm{tr}\left(D(z)2{\sf P}D(z)^{\dag}A\right)\, ,
\end{equation}
In the case of the quantization map $f \mapsto A_f$ based on a weight function $\varpi$, we have
\begin{equation}
\mathfrak{W}_{A_f}(z) = \int_\C \,   \widehat{{\varpi}}(\xi-z)\, f(\xi)\, \frac{\ud^2 \xi}{\pi}\, ,
\end{equation}
which becomes in the case of Weyl-Wigner quantization
\begin{equation}
\label{wigAW}
\mathfrak{W}_{A_f}=f
\end{equation}
(this one-to-one correspondence of the Weyl quantization is related to the isometry property). 

In the case of the anti-normal quantization, the above convolution corresponds to the Husimi transform (when $f$ is the Wigner transform of a quantum pure state).

If the quantization map $f \mapsto A_f$ is regular and isometric, the corresponding inverse map $A \mapsto \mathfrak{W}_A$ is given by
\begin{equation}
\label{invmapquant}
\mathfrak{W}_A = \mathrm{tr} \left(D(z) {\sf M} D(z)^{\dag}A\right)\, , \ \mathrm{where} \ {\sf M} =  {\sf M}^\dag= \int_{\mathbb{C}} \, \varpi(z) D(z)\, \dfrac{\ud^2 z}{\pi} \, . 
\end{equation}
In general this map $A \mapsto \mathfrak{W}_A$ is only the dual of the quantization map $f \mapsto A_f$ in the sense that
\begin{equation}
\int_{\mathbb{C}} \mathfrak{W}_A(z) f(z) \frac{\ud^2z}{\pi}= \mathrm{tr}( A\, A_f)\,.
\end{equation}
This dual map becomes the inverse of the quantization map only in the case of a Hilbertian isometry.

\section{Weyl-Heisenberg integral quantizations of functions and distributions}
\label{sec:fctdist}
\subsection{Acceptable probes $\rho$}
When it is well defined,  the map (\ref{pbformquant}) appears as a  convenient tool to characterize the class of acceptable operator-valued functions ${\sf M}(x)$ and  the class of functions $f$ quantizable with respect to the latter. 
For  the Weyl-Heisenberg integral quantization, we only consider in this section the positive unit trace operators (or ``probe'') ${\sf M} = \rho$  and, in  particular, the examples $\rho= \rho_s$ in the range $\infty < s \leq -1$ including the most manageable CS case. 
Accordingly, the mean value or lower symbol  of $A_f$ is defined by
 \begin{equation}
\label{lowsymbrho}
\check{f}(z) = \int_{\C}  \, \mathrm{tr}\big(\rho(z)\rho(z')\big)\,  f(z')\, \frac{\ud^2z'}{\pi }\, .
\end{equation} 
In particular, the resolution of the identity  proves that:
 \begin{equation}
\label{probrhopos}
 \int_{\C}  \, \mathrm{tr}\big(\rho(z)\rho(z')\big) \frac{\ud^2z'}{\pi } = 1\,,
\end{equation} 
i.e. for each $z$, $\mathrm{tr}\big(\rho(z)\rho(z')\big)=\mathrm{tr}\big(\rho\rho(z-z')\big)$ \underline{is} a probability distribution on the phase space, and so $\check{f}$ is issued from the corresponding  kernel averaging of the original $f$. 
In the CS case $\rho(z)= |z\rg\lg z|$, (\ref{lowsymbrho}) is  the Gaussian convolution (Berezin or heat kernel transform) of the function $f(z)$: 
 \begin{equation}
\label{lowsymb1}
\check{f}(z) = \lg z  | A_f | z \rg = \int_{\C} \, e^{-\left|z-z'\right|^2} f(z')\, \frac{\ud^2z'}{\pi }\, . 
\end{equation}  
An inescapable aspect of any quantization procedure is its classical limit. We should expect that the lower symbol $\check{f}$ of $A_f$ approximates $f$ better and better at this limit. Actually, as is already observed in the Gaussian case, the limit operation is not so straightforward. We first have to give the complex plane a physical phase-space content after introducing physical units  through 
\begin{equation}
\label{physunphsp}
 z \deq \frac{q}{\ell \sqrt{2} } + i \frac{p \ell}{ \hbar \sqrt{2}}\, , 
\end{equation}
 where  $\ell$ is an arbitrary length scale. Secondly, a general operator $\rho$ includes in its definition a set of physical constants, such a temperature, a time scale, etc (like in (\ref{plboltrho})). We thus adopt the following classicality requirement  on the choice of density operators $\rho$
 \begin{defi}
\label{defobscs}
 A density operator $\rho$ is \textit{acceptable from the classical point of view} if
 \begin{enumerate}
  \item[(i)] it obeys the limit condition
  \begin{equation}
\label{rhodirac}
 \mathrm{tr}\big(\rho(z)\rho(z')\big) \to \delta(z- z')\quad \mbox{as} \quad \hbar \to 0\,, \ \ell \to 0\,, \  \hbar/\ell \to 0\, ,
\end{equation}
which  implies a suitable $\hbar$ and $\ell$ dependance on all other parameters involved in the expression of $\rho$, 
  \item[(ii)] the matrix elements $\langle e_n | \rho(z) | e_{n'} \rangle$ (w.r.t. some  orthonormal basis $\{e_n\}$) are $C^\infty$ functions in $z$ with rapid decrease. 
  \end{enumerate}
\end{defi}
Condition (ii) will appear natural for the quantization of distributions.

\subsection{Quantizable functions}

Equation (\ref{lowsymb1}) illustrates nicely the regularizing role of  quantum mechanics  versus classical singularities. 
Note also that the Gaussian convolution allows  to carry out  the semi-classical limit by using a saddle point approximation. For regular functions for which $A_f$ exists, the application of the saddle point approximation is trivial and (\ref{lowsymb1}) holds true. For singular functions, the semi-classical limit is less obvious and has to be verified case by case. Inspired by the CS case in which with mild constraints on $f$ its transform $\check{f}$ inherits infinite differentiability from the Gaussian, let us  adopt  the  second acceptance criterium, which concerns the function $f$ to be quantized.  
\begin{defi}
\label{defobscs}
Given  an acceptable density operator $\rho$, a  function $\C \ni z \mapsto f(z) \in \C$  is  \emph{$\rho$-quantizable} along the  map $f \mapsto A_f$ defined by  (\ref{eqquantvarpi}) with ${\sf M} = \rho$, 
if the map $\C \ni z = \frac{1}{\sqrt{2}} (q + ip) \sim (q,p) \mapsto \check{f}(z)$  is a $C^{\infty}$ function with respect to the $(q,p)$ coordinates of the complex plane.  
\end{defi}

This definition is reasonable insofar as differentiability properties of (\ref{lowsymbrho}) are those of the displacement operator $D(z)$. We will extend this definition to distributions $T \in  \mathcal{D}'(\R^2)$ in the next subsection. 
In the CS case,  the fact that the Berezin transform $f\mapsto \check{f}$ is a Gaussian convolution is of great importance. It explains the robustness of CS quantization, since it is well defined for a very large class of non smooth functions and even, as is shown below, for a  class of distributions including the tempered ones.

\subsubsection*{An example: the quantum angle or phase}

Let us illustrate our approach by revisiting the question of the quantum angle or phase that we have mentioned  in Section \ref{sec:whatquant}. 
Let us write $z = \sqrt J\, e^{i\gamma}$ in action-angle $(J,\gamma)$ notations for the harmonic oscillator \cite{gold81}. The quantization of a function $f(J, \gamma)$ of the action $J\in \R^+$ and  of the  angle $\gamma= \arg(z)\in [0,2\pi)$, which is $2\pi$-periodic in $\gamma$,  yields formally the operator
\begin{equation}
\label{aaquanta}
A_{f} = \int_0^{+\infty}\ud J \int_0^{2\pi}\frac{\ud\gamma}{2\pi} f(J,\gamma)\rho\left(\sqrt{J}e^{i\gamma}\right)\,. 
\end{equation}
 We already mentioned that  the angular covariance property
\begin{equation}
\label{covquantaa}
U_{\mathbb{T}}(\theta)A_f U_{\mathbb{T}}(-\theta)= A_{T(\theta)f}\, , \quad T(\theta)f(J,\gamma) := f(J, \gamma -\theta)\, . 
\end{equation}
holds when the weight function $\varpi$  is isotropic, which implies that $\rho$ is diagonal in the basis $\{ |e_n \rg\}$. 

 In particular, let us quantize with coherent states, $\rho(z)= |z\rg\lg z|$, the discontinuous $2\pi$-periodic angle function $\gimel(\gamma) = \gamma$ for $\gamma \in [0, 2\pi)$. When they are expressed in terms of the action-angle variables  the  standard coherent states  read as  
 \begin{equation}
|z\rg \equiv |J,\gamma\rg = \sum_n \sqrt{p_n(J)} e^{in\gamma} |e_n\rg\, , 
\end{equation}
where $ n \mapsto p_n(J) = e^{-J} J^n/n!$ is the Poisson distribution. The action variable is precisely the Poisson average of the discrete variable $n$, $\lg n\rg_{\mathrm{poisson}} = J$. Note that in electromagnetism, the variables $J$ and $\gamma$  represent  the field intensity (or average number of photons) in suitable units and the phase, respectively. 
 
 Since the angle function is real and bounded,  its quantum counterpart $A_{\gimel}$  is a bounded self-adjoint operator, and it is covariant according \eqref{covquantaa}. In the basis $|e_n\rg$, it is given by  the infinite matrix:
\begin{equation}
\label{scsphaseop}    
A_{\gimel}= \pi\,  1_{{\mathcal H}} + i \, \sum_{n\neq n'}\frac{\Gamma\left( \frac{n + n'}{2}+1\right)}{\sqrt{n!n'!}}\, \frac{1}{n'-n}\, |e_n\rg\lg e_{n'}|\, .
\end{equation}
This operator has spectral measure with support $[0,2\pi]$.  
The corresponding lower symbol reads as the  Fourier sine series:
\begin{align}
\label{lwsymphaseop}
 \lg J,\gamma|A_{\gimel}| J,\gamma\rg &= \pi + i\, e^{-J}\, \sum_{n\neq n'}\frac{\Gamma\left( \frac{n + n'}{2}+1\right)}{n!n'!} \frac{z^{n'}\, \bar{z}^n}{n'-n}\nn \\
 &= \pi -  2\sum_{q = 1}^{\infty}d_q(\sqrt J)\, \frac{\sin{q\gamma}}{q}\, ,
\end{align}
where the function
\begin{align*}
\nonumber d_q (r) &= e^{-r^2}\sum_{m=0}^{\infty} \frac{\Gamma\left( \frac{q}{2}+m+1\right)}{m!(m+q)!}r^{2m+q}\\
&= e^{-r^2}  r^{q}\, \frac{\Gamma(\frac{q}{2} +1)}{\Gamma(q+1)}\, {}_1F_1\left(\frac{q}{2} +1;q+1; r^2\right) 
\end{align*}
balances the trigonometric Fourier coefficient $2/q$ of the angle function $\gimel$. It can be shown 
that this positive function is bounded by 1. 

Let us evaluate the asymptotic behavior of the function  $\lg J,\gamma|A_{\gimel}|J,\gamma\rg$ for small and large $J$ respectively. 
For small $J$, it oscillates around its classical average value $\pi$ with amplitude equal to $\sqrt{\pi J}$:
$$
\lg J,\gamma|A_{\gimel}|J,\gamma\rg \approx \pi - \sqrt{\pi J} \,\sin{\gamma}\,.
$$
As $J \to \infty$, we recover the Fourier series of the $2 \pi$-periodic angle function:
\begin{equation}
\label{largeJ}
\lg J,\gamma|A_{\gimel}|J,\gamma\rg \approx \pi - 2\,\sum_{q = 1}^{\infty}\frac{1}{q}\, \sin{q\gamma} =  \gimel(\gamma) \quad \mbox{for} \quad \gamma \in [0, 2\pi)\, .
\end{equation}
 Such a behavior is understood in terms of the classical limit of 
these quantum objects. Indeed, by re-injecting physical dimensions into our formula, we know that the quantity $\vert z \vert^2 = J$  should appear in the formulas as divided by the Planck constant $\hbar$. Hence, the limit $J\to \infty $ in  our previous expressions can also be considered as the classical limit  $\hbar \to 0$.  Eq. \eqref{largeJ} proves that the behavior of (\ref{lwsymphaseop}) as a function of $\gamma$ for different values of $J$ tends  to the classical behavior as $J \to \infty$. 

We know from \ref{QOintq} that the number operator $\widehat N= \adg \, a$  is, up to a constant shift, the quantization of the classical action, $A_J = \widehat N+1$: 
$A_J = \sum_{n}(n+1)|e_n\rg\lg e_n|$.  
Let us ask to what extent  the commutator of  the action and angle operators and its lower symbol  are close to the  canonical value, namely $i$.
\bea
\label{comactang}
[A_{\gimel},A_J] &= & i \, \sum_{n\neq n'}\frac{\Gamma\left( \frac{n + n'}{2}+1\right)}{\sqrt{n!n'!}}\,  |e_n\rg\lg e_{n'}|\, , 
\\[1mm]
\label{lwscomactang}
\lg J,\gamma|[A_{\gimel},A_J] |J,\gamma\rg &=& 2 i\,  \sum_{q = 1}^{\infty}  d_q(\sqrt J)\, \cos{q\gamma} =: i\, \mathcal{C}(J,\gamma)\, .
\ena
For small $J$, the function $\mathcal{C}(J,\gamma)$ oscillates around 0 with an amplitude equal to $\sqrt{\pi J}$:
$\mathcal{C}(J,\gamma) \approx  \sqrt{\pi J} \,\cos{\gamma}.$
Applying the Poisson summation formula, we get for $J \to \infty$ (or $\hbar\to 0$):
 \begin{equation}
\label{poissonangle}
\lg J,\gamma|[A_{\gimel}, A_J]|J,\gamma\rg \approx -i + 2 \pi i\sum_{n \in \Z} \delta(\gamma - 2 \pi n)\, .
\end{equation} 
One  observes here that, for $\hbar \to 0$, the commutator symbol  becomes  canonical for $\gamma \neq 2 \pi n, \, n \in \Z$. Dirac singularities are located at the discontinuity points of the $2 \pi$ periodic   function   $\gimel(\gamma)$.
The fact that the action-angle commutator is not canonical (see \cite{carnie68} for a comprehensive discussion on this point) should not frighten us  since, on a more general level, we know that there exist such classical canonical pairs for which mathematics (e.g. the Pauli theorem and its correct forms \cite{galapon02}) prevent the corresponding quantum commutator from being exactly canonical).

\subsection{Quantizable distributions}

While proceeding with a quantization scheme, specially with canonical quantization, one usually imposes too  restrictive conditions on the original function $f(z)$. Since it is viewed as a classical observable on the phase space, it is forced to belong to the space of infinitely differentiable functions on $\R^2$, essentially because of the prerequisite Lagrangian and Hamiltonian structures. These conditions prevent from  reaching  a large class of operators, including elementary ones like $\Pi_{n,n'} \deq |e_n\rg\lg e_{n'}|$. Nevertheless, the latter  have also a CS diagonal representation. For that, it is enough to  extend the class of quantizable objects to distributions on $\R^2$ (for canonical coordinates $(q,p)$)  or possibly on $\R^+\times [0, 2 \pi)$ (for  polar coordinates $( r ,\theta)$). 

When we examine the matrix elements of the operator $A_f$ issued from CS quantization, \footnote{The equalities hold in a weak sense, with the already mentioned difficulties for unbounded operators. }
 \begin{align}
\label{csquantmatel1}
\nonumber f \mapsto A_f &= \int_{\C}\,|z\rg\lg z|\, f(z) \, \frac{\ud^2 z}{\pi}\\
&= \sum_{n, n'= 0}^{\infty}  |e_n\rg\lg  e_{n'}| \, \frac{1}{\sqrt{n!n'!}}\, \int_{\C}\, e^{-\vert z \vert^2}z^n{\bar z}^{n'}\, f(z) \, \frac{\ud^2 z}{\pi} \deq \sum_{n, n'= 0}^{\infty}  \left(A_f\right)_{n n'}|e_n\rg\lg  e_{n'}| \, , 
\end{align}
one  immediately thinks of  tempered distributions  on the plane as acceptable objects. Indeed functions  like
\begin{equation}
\label{phinnp}
\phi_{n,n'}(z):=\lg e_n | z\rg \lg z | e_{n'}\rg=\rg e^{-\vert z \vert^2}\, z^n\, \bar{z}^{n'}/\sqrt{n!n'!}
\end{equation}
are rapidly decreasing $C^{\infty}$ functions on the plane with respect to the canonical coordinates $(q,p)$, or equivalently with respect to the coordinates $(z, \bar{z})$: they belong to the Schwartz space $\mathcal{S}(\R^2)$.

Using complex coordinates is  clearly more convenient and we  adopt the following definitions and notations for tempered distributions. First, any function $f(z)$ which is slowly increasing and locally integrable with respect to the Lebesgue measure $\ud^2 z$ on the plane defines a regular tempered distribution $T_f$,  {i.e.,} a continuous linear  form on the  vector space $\mathcal{S}(\R^2)$ equipped with the usual topology of uniform convergence at each order of partial derivatives multiplied by polynomial of arbitrary degree \cite{schwartz61}.
This definition rests on the map,
\begin{equation}
\label{scdist1}
\mathcal{S}(\R^2) \ni \psi \mapsto \langle T_f,\psi \rangle \deq  \int_{\C}\, f(z) \, \psi(z)\, \ud^2z\, .
\end{equation}
For any tempered distribution $T \in \mathcal{S}'(\R^2)$ we define the quantization map $T \mapsto A_T$ as
\begin{equation}
\label{stquantT}
T \mapsto A_T \deq \frac{1}{\pi} \sum_{n, n'= 0}^{\infty}  \langle T,\phi_{n,n'} \rangle |e_n\rg\lg  e_{n'}| \, , 
\end{equation}
where the convergence is assumed to hold in a weak sense. In the sequel the integral notation  will be  kept,  in the usual abusive manner, for all (tempered or not) distributions $T$:
\begin{equation}
\int_{\mathbb{C}} \,|z\rg\lg z|\, T(z) \, \frac{\ud^2 z}{\pi} \deq \frac{1}{\pi} \sum_{n, n'= 0}^{\infty}  \langle T,\phi_{n,n'} \rangle |e_n\rg\lg  e_{n'}|\,.
\end{equation}
In this way, returning to the general scheme (\ref{eqquantvarpi}) with ${\sf M} \equiv \rho$ a positive unit trace operator, we formally define the quantization of a distribution $T$ as
 \begin{equation}
\label{stquantT1}
T \mapsto A_T= \int_{\C}\,  \rho(z) \, \, T(z)\,\frac{\ud^2 z}{\pi} \deq \, \frac{1}{\pi} \sum_{n, n'= 0}^{\infty}  \langle T, \psi_{n,n'} \rangle |e_n\rg\lg  e_{n'}|\,,
\end{equation}
where the 
\begin{equation}
\label{psinnp}
\psi_{n,n'}(z):=\lg e_n | \rho(z) | e_{n'}\rg
\end{equation}
 are assumed to belong to $\mathcal{S}(\R^2)$. The resultant lower symbol is
 \begin{equation}
\label{lowsymbrhodist}
\check{T}(z) = \int_{\C}  \, \mathrm{tr}(\rho(z)\rho(z'))\,  T(z')\, \frac{\ud^2z'}{\pi } = \frac{1}{\pi} \langle \, T,  \textrm{tr} (\rho(z)\rho(\cdot)) \, \rangle\, . 
\end{equation}  
Hence we adopt the following extended definition of quantizable objects.
\begin{defi}
\label{defobscsT}
Given  an acceptable density operator $\rho$, a distribution $T \in \mathcal{D}'(\R^2)$  is  \emph{$\rho$-quantizable}  along the  map $T \mapsto A_T$ defined by (\ref{stquantT1}) if
 the map $\C \ni z = \frac{1}{\sqrt{2}} (q + ip) \sim (q,p) \mapsto \check{T}(z) $  is a smooth ($C^{\infty}$) function with respect to the $(q,p)$ coordinates of the complex plane.
\end{defi}

In the case of CS quantization, we easily check that such definitions are mathematically justified for all tempered distributions. The following result allows  one to extend the set of such acceptable  observables.
\begin{prop}
\label{defobsdist}
A distribution   $T \in \mathcal{D}'(\R^2)$ is   \emph{CS quantizable}
if there exists $\eta < 1$ such that the product $e^{- \eta \vert z \vert^2}\, T \in \mathcal{S}'(\R^2)$, i.e. is a tempered distribution.
\end{prop}
Note that  extensions to distributions  have been considered in  \cite{-sudarshan1}, \cite{-gros_par}, and \cite{boggiatto03,boggiatto04} for the Weyl-Wigner quantization. 
In the general $\rho$ case, one expects to have a similar result with suitably chosen weight functions $\varpi(z)$. In the sequel, we suppose that such a choice has been made. 

Thus, according to Proposition \ref{defobsdist},  the definition \ref{defobscs} can be extended to locally integrable functions $f(z)$   increasing like $e^{\eta \vert z \vert^2}\, p(z)$ for some $\eta <1$ and some polynomial $p$. It becomes straightforward to generalize  to distributions. We know that  the latter are characterized as derivatives (in the distributional sense) of such functions.  We recall here that  partial derivatives of distributions  are given by
\begin{equation}
\label{scdist21}
 \llg \frac{\partial^r}{\partial z^r}\, \frac{\partial^s}{\partial \bar{z}^s}\,T,\psi \rrg = (-1)^{r+s}\, \llg T, \frac{\partial^r}{\partial z^r}\, \frac{\partial^s}{\partial \bar{z}^s}\,\psi \rrg \, .
\end{equation}
We also recall that the multiplication of distributions $T$ by smooth functions $\alpha(z) \in C^{\infty}(\R^2) $ is understood through: 
\begin{equation}
\label{prodscdist}
C^{\infty}(\R^2) \ni \psi \mapsto \langle \alpha T,\psi \rangle := \langle  T,\alpha\, \psi \rangle\, .
\end{equation}
Of course, all compactly supported distributions like Dirac's and its derivatives,  are tempered and so are expected to be $\rho$-quantizable. The Dirac distribution supported by the origin of the complex plane is  denoted as usual by $\delta$ (and in the present context by $\delta(z)$) :
\begin{equation}
\label{dirac1}
C^{\infty}(\R^2) \ni \psi \mapsto \langle \delta,\psi \rangle = \int_{\C}\, \delta (z) \, \psi(z)\, \ud^2z \deq \psi(0)\,. 
\end{equation}
As a first example, let us  $\rho$-quantize the Dirac distribution.
\begin{equation}
\int_{\C} \rho(z) \, \pi \delta( z)\, \frac{\ud^2 z}{\pi} =  \rho(0) \equiv \rho\, .
\label{quantdirrho}
\end{equation}
In particular, in the CS case, we find that the ground state projector is the quantized version of the Dirac distribution supported at the origin of the phase space.
\begin{equation}
A_{\pi \delta} =  | e_0\rangle \langle e_0| \, .
\label{quantdir}
\end{equation}
Similarly, the quantization of the Dirac distribution $\delta_{z_0} \equiv \delta(z-z_0)$ at the point $z_0$ yields the displaced density matrix:
\begin{equation}
\label{quantdz}
A_{\pi \delta_{z_0}} = D(z_0) \rho  D^\dag(z_0) = \rho(z_0)\, .
\end{equation}
In the CS case, we find the CS projector with parameter $z_0$:
\begin{equation}
\label{quantdz}
A_{\pi \delta_{z_0}} = D(z_0) | e_0\rangle \langle e_0|  D^\dag(z_0) = | z_0\rangle \langle z_0|\, .
\end{equation}
Thus, the density matrix $\rho$, which is, besides the measure $\nu$, the main ingredient of our quantization procedure is precisely the quantized version of the Dirac distribution supported at the origin of the phase space. We have here the key for understanding the deep meaning of this type of quantization: replacing the classical states $\delta_{z_0}$, i.e. the highly abstract points of the phase space, physically unattainable, by a more realistic object, $\rho(z_0)$, a kind of ``inverted glasses'' chosen by us, whose the probabilistic content takes into account the measurement limitations of any localization apparatus. The operator $\rho$ can be viewed as a probe whose the displaced versions give a quantum portrait of the euclidean plane. 

 The obtention of all possible  projections $\Pi_{nn} = |e_n\rg\lg e_n |$ or even all possible simple operators $\Pi_{nn'} = |e_n\rg\lg e_{n'}|$ is based on the quantization of  partial derivatives of the $\delta$ distribution. 

\subsection{About inversion formulae}

From now on, we restrict our study to CS quantization which yields easily explicit formulas in many cases of interest. Thus the quantized  versions derivatives of the Dirac distribution read as: 
\begin{equation}
A_{\pi \partial_z^n \partial_{\bar{z}}^{n'} \delta} = (-1)^{n+n'} n!\, n'!\, \sum_{p=0}^{\inf{(n,n')}} \frac{(-1)^p}{p!} \frac{1}{\sqrt{(n-p)! \, (n'-p)!}} \Pi_{n-p,n'-p}\,.
\label{quantdirderpow}
\end{equation}

With this quantity $A_{\pi \partial_z^n \partial_{\bar{z}}^{n'} \delta}$ at hand, one can  invert the formula \eqref{quantdirderpow} in order to get $\Pi_{n,n'}=\left|e_n\right\rangle\left\langle e_{n'}\right|$  as:
\begin{equation}
\label{oblproj}
\Pi_{n,n'}=(-1)^{n+n'} \sqrt{n! \, n'!} \sum_{p=0}^{\inf (n,n')} \frac{1}{p! \, (n-p)! \, (n'-p)!}  A_{\pi \partial_z^{n-p} \partial_{\bar{z}}^{n'-p} \delta} \, \,.
\end{equation}
Therefore the operator $\Pi_{n,n'}$ is the CS quantization  of the distribution $T_{n,n'}$ supported at the origin:
\begin{equation}
\label{upsymbobl}
T_{n,n'} = (-1)^{n+n'} \sqrt{n! \,n'!} \sum_{p=0}^{\inf (n,n')} \frac{1}{p! (n-p)! (n'-p)!} \, \left\lbrack\frac{\partial^{n-p}}{\partial z^{n-p}}\, \frac{\partial^{n'-p}}{\partial \bar{z}^{n'-p}}\delta \right\rbrack\, .
\end{equation}
Let us note that this distribution, as  is well known, can be approached, in the sense of the  topology on $\mathcal{D}'(\R^2)$ (or $\mathcal{S}'(\R^2)$), by smooth functions, such as linear combinations of derivatives of Gaussians.  
The projectors $ \Pi_{n,n}$ are then obtained trivially from  \eqref{oblproj} to get
\begin{equation}
\label{diagproj}
| e_n \rangle \langle e_n| = \Pi_{n,n}=\sum_{p=0}^n \frac{1}{p!} \left( \begin{array}{c} n \\ p \end{array} \right) A_{\pi \partial_z^p \partial_{\bar{z}}^p \delta}\, .
\end{equation}
A warning is in order at this point. Using formal manipulations, one can be tempted to use \eqref{upsymbobl} to obtain the classical counterpart (or upper symbol) of any projector $| \psi \rangle \langle \psi|$ as
\begin{equation}
\label{formalupperproj}
| \psi \rangle \langle \psi| = A_T \quad \textrm{with} \quad T = \sum_{n,n'=0}^\infty \langle e_n|\psi \rangle \langle \psi | e_{n'} \rangle \, T_{n,n'} \,.
\end{equation}
Unfortunately, the infinite series expansion involved to define $T$ does not converge in general in  $\mathcal{S}'(\mathbb{R}^2)$ (or  $\mathcal{D}'(\mathbb{R}^2)$), and therefore the previous expression has no mathematical meaning. Only for finite linear combinations, one can assert the existence of $T$ without studying the convergence problems. As an illustration of these problems, we choose $| \psi \rangle =| z_0 \rangle$. Using \eqref{quantdz}, \eqref{formalupperproj} and the fact that the map $T \mapsto A_T$ is injective on  $\mathcal{S}'(\mathbb{R}^2)$ (this results from the density of finite linear combinations of the $\phi_{n,n'}$ in $\mathcal{S}(\mathbb{R}^2)$) we should have:
\begin{equation}
\delta_{z_0} = e^{-|z_0|^2} \sum_{n,n'=0}^\infty \frac{z_0^n \bar{z}_0^{n'}}{\sqrt{n! \, n'!}} T_{n,n'}\,.
\end{equation}
But all the $T_{n,n'}$ are supported at the origin, then if the series converges in $\mathcal{S}'(\mathbb{R}^2)$, it is also supported at the origin, while $\delta_{z_0}$ is supported at the point $z_0$. Therefore the previous equality is wrong: in fact the series involved in the r.h.s does not converge in $\mathcal{S}'(\mathbb{R}^2)$.
This example illustrates the difficulties of formal manipulations in order to construct  an inversion (dequantization) operator $\mathcal{I}_Q$. Its mathematical existence is in fact  restricted: from above it is only proved for operators $B$ which are finite rank w.r.t. the basis $\{|e_n\rg\}$.  $\mathcal{I}_Q$ is  given by:
\begin{equation}
\label{inversion}
\mathcal{I}_Q(B)=\sum_{n,n'} \left\langle e_n \right| B \left| e_{n'} \right\rangle T_{n,n'}
\end{equation}

When is properly defined, this inversion map also enables us to construct  a star product $*$ on the classical phase space verifying $A_{f\star g}=A_f \: A_g$ (see for instance \cite{hirshfeld02} for a general review on deformation quantization, and \cite{voros02,daoud03,alex01,bal07} for more  material based on coherent states)
$$
f\star g=\mathcal{I}_Q \left(A_f \: A_g\right)\, . 
$$
Note that this star product involves the original $f,g$ (i.e., upper symbols) in contrast to the Voros star product \cite{voros02} which involves the lower symbols.

Many of the  ideas  around this combination of coherent states with distributions pertain to the domain of Quantum Optics (Q.O.). They are already present in the original works by Sudarshan \cite{-sudarshan1}, Glauber \cite{-glauber1},  Klauder \cite{-klausud}, Cahill \cite{cahill65}, Miller \cite{miller68} and others. In Quantum Optics the basic idea is that replacing the non diagonal representation of quantum operators, say $T$, (usually, in this context,  one focuses on  the  density operators $\rho$) given by 
$$
T=\int_{\C^2} \left|z_1\rangle\langle z_2\right|\, \, \left\langle z_1\right|A\left|z_2\right\rangle \, \ud^2z_1\ud^2z_2\,  
$$
by a diagonal one, also called the $P$-representation,
$T=\int_{\C}\, |z\rangle\langle z|\, P(z, \bar z) \,  \ud^2z$, can  simplify considerably some calculations. Although this can be considered as the CS quantization of $P(z)$, the spirit is quite different since the Q.O. approach is the inverse of the one presented here: given an $A$, the  question is to find $P(z)$. The main result   obtained  in this direction is that one can formally write a $P$-representation for each quantum operator $T$, which is  given by \cite{-sudarshan1}
\begin{equation}
\label{sudformula}
P(z=r e^{i \theta})=\sum_{m,n=0}^\infty \frac{\left\langle n\right|T\left|m\right\rangle \sqrt{n! m!}}{2 \pi r (n+m)!}  e^{r^2+i (m-n) (\theta-\pi)}\, 
\delta^{(m+n)}(r)
\end{equation}
or by  \cite{kl}
\begin{equation}
\label{klformula}
 P\big(z=(q+ip)/\sqrt{2}\big)=\mathcal{F}^{-1} \Big[ \tilde{T}(x,y) \;e^{\frac{x^2+y^2}{2}}\Big] \, ,
\end{equation}
where $\tilde{T}(x,y)=\mathcal{F} \left[ \left\langle z\right|T\left|z\right\rangle \right].$
Here $\mathcal{F}$  is the Fourier transform from the $(q,p)$-space  to the $(x,y)$-space, and $\mathcal{F}^{-1}$ is its inverse.
However the question of the validity of such formulas is mathematically nontrivial: the convergence in the sense of distributions of (\ref{inversion},\ref{sudformula}) is a difficult problem (which might even have no explicit solution!), and  for instance has  been partially studied  by Miller in \cite{miller68}.
Manifestly,  the works done in this direction were concentrated on the dequantization problem (finding an associated classical  function to each quantum operator) and this was done in a quite pragmatic spirit in order to simplify computations. Let us note that the existence of such a well-defined dequantization procedure is by no means a physical requirement, since the quantum realm is by definition richer than the classical one. A more physical requirement is that the semi-classical limit be well behaved, as we stressed in Section \ref{sec:whatquant}. 

\section{Affine or wavelet quantization}
\label{sec:affine}

We present here a basic and deep illustration of the content of Subsection \ref{subsec:covq}.  The measure set $X$ is the upper half-plane $ \{(q,p)\, |\, p\in \mathbb{R}\, , \, q> 0\}$. For a sake of simplicity, we suppose that variables $q$ and $p$ are physically dimensionless (physical dimensions are easily restored through injection of appropriate scaling factors into the formalism). 
 Equipped with the multiplication rule
 \begin{equation}
\label{multiaff}
(q,p)(q_0,p_0)=\left(qq_0, \frac{p_0}{q}+p\right), ~q\in \mathbb{R}^{\ast}_+, ~p\in \mathbb{R}
\end{equation}
$X$ is viewed as the  affine group Aff$_+(\R)$ of the real line, acting as $x \mapsto (q,p)\cdot x = p +qx$.  Group  Aff$_+(\R)$ has
two non-equivalent  UIR, $U_{\pm}$ \cite{gelnai,aslaklauder}. Both are square integrable and they are  crucial ingredient of   continuous wavelet analysis \cite{-grosmor,-gros1,-gros2,aag00}. 
Representation $U_+\equiv U$ is carried on by  the Hilbert space $\mathcal{H} = L^2(\mathbb{R}^{\ast}_+,  \ud x)$:
\begin{equation}
\label{U+}
U(q, p) \psi(x) = \frac{e^{i px}}{\sqrt{q}} \psi\left( \frac{x}{q}\right)\, .
\end{equation}
Picking a unit-norm state $\psi \in L^2(\mathbb{R}_+^\dagger, \ud x)\cap L^2(\mathbb{R}_+^\dagger, \ud x/x)$ (named fiducial vector or mother wavelet in the context of wavelet analysis) produces all affine  coherent states (or wavelets)  defined as in \eqref{Gcs}:
\begin{equation}
\label{affwave}
| q, p \rangle\ = U(q,p) | \psi \rangle\, . 
\end{equation}
Square integrability of the UIR $U_+$ and admissibility yield the resolution of the identity
\begin{equation}
\label{resunaff}
\int_{\Pi_+} \dfrac{\ud q \ud p}{2 \pi  c_{-1}} | q, p \rangle \langle q, p | = I \, , \ \mbox{where}\  
c_{\gamma}:=\int_0^\infty\frac{\ud x}{x^{2+\gamma}}\, \vert\psi(x)\vert^2\, .
\end{equation}
The covariant integral quantization follows: 
\begin{equation}
\label{affcovquant}
f \ \mapsto \ A_f = \int_{\Pi_+} \dfrac{\ud q \ud p}{2\pi  c_{-1}}  f(q, p) |q, p \rangle \langle q, p |
\end{equation}

Note that the idea of proceeding in quantum  theory with an ``affine'' quantization instead of the  Weyl-Heisenberg quantization  was already present in Klauder's  work \cite{klauder_Aslaksen70} devoted the question of singularities in quantum gravity (see \cite{klauder11} for recent references). The procedure followed by Klauder, which we summarize in Appendix \ref{affqgr}, rests on the representation of the affine Lie algebra. In this sense, it remains closer to the canonical one and it is not of the integral type. Also note that we have restricted  the general construction  encapsulated by \eqref{intgrR}  to the projector $\sf{M}= |\psi\rg\lg\psi|$. It is certainly appealing to examine, as we did for the Weyl-Heisenberg group, quantizations based on more general operators like
\begin{equation}
\label{affright}
{\sf M}= \int_{\Pi_+} U(q,p)\, \varpi(q,p)\, \ud q\ud p\, , 
\end{equation}
where $\varpi(q,p)$ is a suitably chosen weight function on the half-plane.

The first interesting issue of the map \eqref{affcovquant} is that the quantization yields canonical commutation rule, up to a scaling factor,  for $q$ and $p$:
\begin{equation}
\label{affqcan}
A_p= P = -i\frac{d}{d x}\, , A_{q} =({c}_{0}/c_{-1})\, Q\, , \quad Qf(x) :=  x f(x)\,, \quad [A_q,A_p]=({c}_{0}/c_{-1}) iI\, . 
\end{equation}
However, if on one hand $Q$ is (essentially) self-adjoint, on the other hand we know from \cite{reedsimon2} that $P$ is symmetric but   has no self-adjoint extension. 
The quantization of any power of $q$ for which the integral \eqref{affcovquant} is well defined is, up to a scaling factor, canonical:
\begin{equation}
\label{affqpow}
A_{q^\beta} =({c}_{\beta-1}/c_{-1})\, Q^\beta\, .
\end{equation}
A second striking result issue of the procedure is a  regularization of the quantum ``kinetic energy'': 
 \begin{equation}
\label{kinquantaff}
 A_{p^2} = P^2 +  KQ^{-2}\  \mbox{with}\   K=K(\psi)= \int_0^{\infty}\frac{\ud u}{c_{-1}}\,u \, (\psi'(u))^2\,, 
\end{equation}
i.e., the quantization procedure always yields an additional term. This term is a centrifugal potential whose the strength depends on the fiducial vector only, and can be made as small as one wishes through an appropriate choice of $\psi$. In other words, this affine or wavelet quantization forbids a quantum free particle moving on the positive line to reach the origin, and the centrifugal effect could be very tiny out a region near the origin, unattainable to any realistic measurement, as it is the case in microscopic physics with central potentials.  
Now, if we consider the quantum dynamics of  such a free motion, it is known \cite{reedsimon2,Gesztesy} that  the operator $P^2= -d^2/dx^2$ alone  in $L^2(\mathbb{R}^{\ast}_+,  \ud x)$ is not essentially self-adjoint whereas the regularized operator \eqref{kinquantaff} is for $K \geq 3/4$.  It follows that for $K \geq 3/4$ the quantum dynamics is unitary during the entire evolution, in particular in the passage from the motion  towards $Q = 0$ to the motion away from $Q = 0$. 
For instance, one can choose as 
a fiducial vector the smooth function on $\R^\ast_+$ with parameters $a>0$ and  $b>0$
\begin{equation*}
\psi_{\nu}(x) = c_{a,b}e^{-(a/x + bx)}\, , 
\end{equation*}
where $c_{a,b}$ is a normalization factor. This function is deduced from the well-known prototype of test functions in $\mathcal{D}(\R)$:
\begin{equation*}
\omega_{\alpha}(t) = \left\lbrace\begin{array}{cc}
   \exp\left(-\frac{\alpha}{1-t^2}\right)   &  -1<t<1  \\
    0  &   \vert t\vert \geq 1 
\end{array}\right.\, , \quad \alpha >0\, , 
\end{equation*}
through the conformal transform $t= (\mu x -1/2)/(\mu x + 1/2)$, $\mu =\sqrt{b/a}/2$, $\alpha = 4 \sqrt{ab}$, which maps the bounded interval $(-1,1)$ onto the unbounded $(0, +\infty)$. This function allows to go easily through analytical calculations whose  the details are given in   \cite{wcosmo1} where the present  procedure is applied to quantum gravity. 

\section{Conclusion}
 \label{sec:conclu}
 The primary conclusion we draw from the approach and material presented in this paper is that integral quantization(s) pave(s) the way to a considerable  realm of freedom for investigating  the relation between classical and quantum representations of mathematical models. Of course, from physicist's viewpoint, the unique criterium of validity of one or a class of choices made among so many possibilities offered by the formalism is consistent with measurement issues.  Now, the advantages of the method with regard to other quantization procedures in use are of four  types. 
 
 \begin{enumerate}
  \item[(i)] The minimal amount of constraints imposed to the classical objects to be quantized. 
  \item[(ii)] Once a choice of (positive) operator-valued measure has been made, there is no ambiguity in the issue, contrarily to other method(s) in use. We think in particular to the ordering problem hampering canonical quantization. To one classical object corresponds one and only one quantum object provided that the integral \eqref{formquant} makes sense. 
  \item[(iii)] The method produces in essence a regularizing effect, at the exception of certain choices,  like the Weyl-Wigner integral quantization. 
  \item[(iv)] The method, through POVM choices, offers the possibility to keep a full probabilistic content. As a matter of fact,  the Weyl-Wigner integral quantization does not rest on a POVM. 
\end{enumerate}
 
 Now  one can argue about that question of freedom in choosing the operator-valued measure on which is built a given integral quantization. On the contrary, we think that this exemplifies the strength of the method. We here appeal to the methods in use in signal analysis, the domain we start with in our introduction. Given a signal, an image, ..., signalists have been developing for many years, mostly during the past 25 years, a huge variety of tools departing from the traditional Fourier analysis: Gabor, wavelet, ridgelet, ....\cite{aag00}. Each one of these various tools favors particular aspects of the signal: frequency, time-frequency, time-scale ... All these tools are complementary, and the only exigences are the computer memory and rapidity resources and the absence of some inopportune artifact. It is remarkable that all these approaches are worked out in terms of integral transforms involving (positive or not) operator valued measures, the most elementary being Fourier, the counterpart, in a certain sense, of the Weyl-Wigner quantization. Gabor analysis is the counterpart of the standard CS  quantization, and wavelet analysis is the the counterpart of affine integral quantization. These mutual irrigations between quantum physics and signal analysis deserve a lot more attention in future  investigations. 

\appendix
\section{Weyl-Heisenberg algebra, group, and displacement operator: a glossary}
\label{glossDz}
\subsection*{Weyl-Heisenberg algebra and its Fock or number representation}
\begin{itemize}
\item Notational convention: set of nonnegative integers is $\NN= 0,1, 2 , \dotsc$.
\item Let $\mathcal{H}$ be a separable (complex) Hilbert space  with orthonormal basis $e_0,e_1,\dots, e_n \equiv |e_n \rangle, \dots$, (e.g the Fock space with $|e_n\rg \equiv |n\rg$. 

\item Define the lowering and raising operators $a$ and $a^\dag$ as
\begin{align}
   a\, \ket{e_n} & = \sqrt{n} \ket{e_{n-1}}\, , \quad  a\ket{e_0} = 0 \, \quad \mbox{(lowering operator)}\\
   a^{\dag} \, \ket{e_n} & = \sqrt{n +1} \ket{e_{n+1}} \quad \mbox{(raising operator)}\, .
\end{align}  
\item Equivalently
\begin{align}
   a\,  & =\sum_{n=0}^\infty \sqrt{n+1} \ket{e_n} \bra{e_{n+1}}  \quad \mbox{(weak sense)}\\
   a^{\dag} \,  & = \sum_{n=0}^\infty \sqrt{n+1} \ket{e_{n+1}} \bra{e_{n}}  \quad \mbox{(weak sense)}\, . 
\end{align}  
  \item Operator algebra $\{a,\adg, I\}$ is defined by
  \begin{equation}
\label{ccr}
  [a,\adg]= \mbox{\large 1}\,.
\end{equation}
  \item Number operator: $N= \adg a$, spectrum $\NN$, $N|e_n\rg = n |e_n\rg$.  
\end{itemize}

\subsection*{Unitary Weyl-Heisenberg group representation }
\begin{itemize}
\item To each complex number $z$ is associated the (unitary) displacement operator  or ``function $D(z)$'' :
\begin{equation}
\label{displac}
\C \ni z \mapsto D(z) = e^{z\adg -\bar z a}\, ,\quad D(-z) = (D(z))^{-1} = D(z)^{\dag}\, . 
\end{equation}
\item Using the Baker-Campbell-Hausdorff formula we have
\begin{equation}
\label{camhaus}
D(z)=e^{z a^\dag} e^{-\bar{z} a} e^{-\frac{1}{2} |z|^2}=e^{-\bar{z} a} e^{z a^\dag}e^{\frac{1}{2} |z|^2},
\end{equation}
\item It follows the formulae:
\begin{equation}
\label{propdisp1}
\dfrac{\partial}{\partial z} D(z) = \left(a^\dag - \dfrac{1}{2} \bar{z} \right) D(z) = D(z) \left( a^\dag + \dfrac{1}{2} \bar{z} \right).
\end{equation}
\begin{equation}
\label{propdisp2}
\dfrac{\partial}{\partial \bar{z}} D(z) = - \left(a - \dfrac{1}{2} z \right) D(z) = - D(z) \left( a + \dfrac{1}{2} z\right).
\end{equation}
\item Addition formula:
\begin{equation}
\label{additiveD}
D(z)D(z') = e^{\frac{1}{2}z \circ z'}D(z+z') \,,
\end{equation}
where $z \circ z'$ is the symplectic product $z \circ z'= z\bar{z'} -\bar{z} z'= 2i\mathrm{Im} (z\bar z') = - z' \circ z$. 
\item It follows the covariance formula on a global level:
\begin{equation}
\label{Dcovglob}
D(z) D(z') D(z)^\dag = e^{z \circ z'} D(z').
\end{equation}
\item and on a Lie algebra level
\begin{equation}
\label{Dcovinf}
 \quad D(z) a D(z)^\dag= a - z, \quad D(z) a^\dag D(z)^\dag= a^\dag - \bar{z}, 
\end{equation}
\item Matrix elements of operator $D(z)$ involve associated Laguerre polynomials $L^{(\alpha)}_n(t)$:
\begin{equation}
\label{matelD}
\lg e_m|D(z)|e_n\rg = D_{m n}(z) =  \sqrt{\dfrac{n!}{m!}}\,e^{-\vert z\vert^{2}/2}\,z^{m-n} \, L_n^{(m-n)}(\vert z\vert^{2})\, ,   \quad \mbox{for} \ m\geq n\, , 
\end{equation}
with $L_n^{(m-n)}(t) = \frac{m!}{n!} (-t)^{n-m}L_m^{(n-m)}(t)$ for $n\geq m$.
  \item Weyl-Heisenberg group: 
  \begin{align}
\label{weylheisG}
\nonumber G_{\mathrm{WH}}&= \{(s,z)\, , \, s\in \R , \, z\in \C\}\, \\ (s,z)(s',z') & = (s+s' + \mathrm{Im} (z\bar z'), z+z')\, ,\quad (s,z)^{-1} = (-s,-z)\, . 
\end{align}
  \item Unitary representation by operators on $\mathcal{H}$ (consistent with (\ref{additiveD}) and (\ref{weylheisG})): 
  \begin{equation}
\label{unrepWH}
(s,z) \mapsto e^{is}D(z)\, , \quad (s,z)(s',z') \mapsto e^{is}D(z)e^{is'}D(z') = e^{i(s+s' + \Im z\bar z')} D(z+z')\,. 
\end{equation}
\end{itemize}

\subsection*{Parity and time reversal}
To complete the picture one  defines  two discrete symmetries.
\begin{itemize}
  \item The first one is the ``parity" ${\sf P}$ acting on $\mathcal{H}$ as a linear operator through
\begin{equation}
\label{parity}
{\sf P} \ket{e_n}= (-1)^n \ket{e_n}\, , \quad \mbox{or}\quad {\sf P}= e^{i \pi a^\dag a}\,. 
\end{equation}
 \item The second symmetry is the so-called ``time reversal" ${\sf T}$ acting on $\mathcal{H}$ as a conjugation, that is an {\it antilinear operator} such that
\begin{equation}
\label{timerev}
{\sf T} \ket{e_n}= \ket{e_n}\, .
\end{equation}
 \item These discrete symmetries verify
\begin{align}
 {\sf P}^2 &={\sf T}^2 =I, \\
 {\sf P} a {\sf P}& =-a \;; {\sf P} a^\dag {\sf P} = -a^\dag,\\
  {\sf T} a {\sf T}& =a \;; {\sf T} a^\dag {\sf T} = a^\dag,\\
  {\sf P} D(z) {\sf P}&=D(- z) \;; {\sf T} D(z) {\sf T} = D(\bar{z}).
 \end{align}
\end{itemize}

\subsection*{Rotation in the plane}
 The unitary representation $\theta \mapsto U_{\mathbb{T}}(\theta)$ of the torus $\SN^1$ on the Hilbert space $\mathcal{H}$ is defined as 
\begin{equation}
\label{unrotplane}
U_{\mathbb{T}}(\theta)|e_n\rg = e^{i (n + \nu) \theta}|e_n\rg\, , \quad \nu \in \R\, .
\end{equation}
Note that ${\sf P}= U_{\mathbb{T}}(\pi)$ with $\nu = 0$. 
 We then obtain the  rotational covariance of the displacement operator.
\begin{equation}
\label{rotcovD}
U_{\mathbb{T}}(\theta)D(z)U_{\mathbb{T}}(\theta)^{\dag} = D\left(e^{i\theta}z\right)\, . 
\end{equation}

\subsection*{Integral formulae for $D(z)$}
\begin{itemize}

\item A first  fundamental integral: from
\begin{equation}
\label{fundintD}
\int_0^{\infty}dt\, e^{-\frac{t}{2}}\, L_n(t)  = (-2)^n\, \Rightarrow \int_{\C} \frac{d^2 z}{\pi}\, D_{m n}(z)= \delta_{mn} (-2)^m\,,
\end{equation}
it follows 
\begin{equation}
\label{fourD0}
\int_{\C} \frac{d^2 z}{\pi}\, D(z) = 2{\sf P}\, ,
\end{equation}
\item A second fundamental integral: from (\ref{matelD}) and  the orthogonally of the associated Laguerre polynomials we obtain the ``ground state'' projector $P_0$ as the Gaussian average of $D(z)$:
\begin{equation}
\label{lapD}
 \int_{\mathbb{C}} \dfrac{d^2 z}{\pi} e^{- \frac{1}{2} |z|^2} D(z)= \ket{e_0} \bra{e_0} \,. 
\end{equation}
\item More generally for $\Re(s) < 1$
\begin{equation}
\label{lapDs}
\int_{\mathbb{C}} \dfrac{d^2 z}{\pi} e^{\frac{s}{2} |z|^2} D(z) = \dfrac{2}{1-s} \exp \left( \ln \dfrac{s+1}{s-1} a^\dag a \right)\,, 
\end{equation}
where the convergence holds in norm for $\Re(s)<0$ and weakly for $0 \leq \Re(s) < 1$.
 \end{itemize}

\subsection*{Harmonic analysis on $\C$ and symbol calculus}
\begin{itemize}
  \item ``Symplectic Fourier transform" on $\mathbb{C}$:
let $f$ be a $L^1$ function on $\mathbb{C}$, its symplectic transform $\hat{f}$ is defined as:
\begin{equation}
\label{symfourier}
\hat{f}(z)=\int_{\mathbb{C}} \dfrac{d^2 \xi}{\pi} e^{ z \circ \xi } f(\xi)\, .
\end{equation}
  \item The symplectic transform is an involution ($\hat{\hat{f}}=f$). 
 \item The usual symbolic integral calculus yields the Dirac-Fourier formula:
\begin{equation}
\label{diracfourier}
\int_{\mathbb{C}} \dfrac{d^2 \xi}{\pi} e^{z \circ \xi }= \pi \delta^{(2)}(z)
\end{equation}
\item The  resolution of the identity  follows from (\ref{fourD0}):
\begin{equation}
\label{resunweyl}
\int_{\C} \frac{d^2 z}{\pi}  \, D(z)\, 2{\sf P}\, D(-z) = I\, . 
\end{equation}
This formula is at the basis of the Weyl-Wigner quantization (in complex notations)
 \item The Fourier transform of operator $D$, is easily found from (\ref{fourD0}) and Fourier transform of the addition formula (\ref{additiveD}): 
  \begin{equation}
\label{ftransfD}
\int_{\C} \frac{d^2 z'}{\pi} \,e^{z \circ z'} \, D(z') = 2 D(2z){\sf P} = 2 {\sf P} D(-2z)  \,.
\end{equation}
\end{itemize}

\section{ Klauder's affine quantization}
\label{affqgr}

One starts from the classical  ``$ax + b$'' affine algebra with its two generators $q$ (position), $d=pq$ (dilation), built from the usual phase space canonical pair $(q,p)$, $\{q,p\}=1$, and obeying
$\{q,d\}= q$. Then, following the usual canonical quantization procedure, 
$q\mapsto Q$, $p\mapsto P$, with $[Q,P]= i\hbar I$,  one obtains the quantum version of the dilation, $d\mapsto \frac{1}{2} (PQ + QP)$, and the resulting affine commutation rule $[Q,D]= i\hbar Q$ for these self-adjoint operators. At the difference of the original $Q$ and $P$, the affine operators $Q$ and  $D$ are reducible: there are three inequivalent irreducible self-adjoint representations, $Q>0$, $Q<0$, and $Q=0$.   The quantization of classical observables follows through the usual replacement $f(q,d) \mapsto f(Q,D)$ followed by a symmetrization. Then, a specific family of  affine coherent states $|p,q\rg$ (in Klauder's notation) is built from the unitary action of the affine group
\begin{equation}
\label{affcsfid}
|p,q\rg := e^{ipQ/\hbar}e^{-i \ln(q) D/\hbar}|\tilde\beta\rg\, ,
\end{equation}
on a fiducial vector $|\tilde\beta\rg $ chosen as an extremal weight vector which is a solution of the  first order differential  equation 
\begin{equation}
\label{eqdifffid}
(Q-1 + (i/\tilde\beta)D)|\tilde\beta\rg= 0,
\end{equation}
where $\tilde\beta$ is a free parameter. Note that this equation is the affine counterpart of the $a|0\rg =0$  satisfied by the Gaussian fiducial vector in the case of standard coherent states. 

Given a quantum operator $A$ issued from this scheme, like the Hamiltonian, its mean values or lower symbols $A(p,q) = \lg p,q |A |p,q\rg$ allows to make the classical and quantum theories coexist in a consistent way:  the classical limit of this \textit{enhanced affine quantization \`a la} Klauder is a canonical theory. 

In  \cite{klauder_Aslaksen70} the authors build  a toy model of gravity where  $p>0$ represents the metric with signature constraints and  $q$  represents the Christoffel symbol. In the later work,  Klauder has chosen  $q>0$  for the metric and $ -p$  as the Christoffel symbol.

In a recent paper \cite{fanuel_Zonetti13}, Fanuel and Zonetti follow Klauder's approach to affine quantization to deal with highly symmetric cosmological models.  

J. R. Klauder has also studied affine quantization of the entire gravitational field. A short and summarizing article is  \cite{klauder11} with references therein.


\begin{thebibliography}{99}


 \bibitem{kiukas} J. Kiukas, P. Lahti, and K. Ylinenc, Phase space quantization and the operator moment problem,
  \textit{J. Math. Phys. }  \textbf{47} (2006) 072104 
  
  \bibitem{weyl28}  
{ H. Weyl, \textit{Gruppentheorie und Quantenmechanik} (Hirzel, Leipzig,1928); 
H. Weyl, \textit{The Theory of Groups and  Quantum Mechanics} (Dover, New York, 1931)}

\bibitem{grossmann76}  A. Grossmann, Parity operator and quantization of $\delta$-functions, 
          \textit{Commun.~Math. Phys.}  \textbf{48} (1976)  191--194

\bibitem{daub11}I.~Daubechies, On the distributions corresponding to bounded operators  in the Weyl quantization,  
\textit{Commun. Math. Phys.} \textbf{75}   229--238 (1980)

\bibitem{daub12} I.~Daubechies and A.~Grossmann, An integral transform related to quantization. I. 
        \textit{J. Math. Phys.} \textbf{21}   2080--2090 (1980)

\bibitem{daub13} I.~Daubechies, A.~Grossmann,  and J.~Reignier, An integral transform related to  quantization.  II.
       \textit{J. Math. Phys.}  \textbf{24}  239--254 (1983)

  
  \bibitem{zachos}  C. Zachos, D. Fairlie, and T. Curtright   2006 \textit{Quantum Mechanics in Phase Space: An Overview With Selected Papers}, (Singapore: World Scientific Publishing, 2006)
  
 
\bibitem{herzberg1989} G. Herzberg,  \textit{Molecular Spectra and Molecular Structure:
Spectra of Diatomic Molecules, 2nd. ed.} (Krieger Pub., Malabar, FL,  1989)

\bibitem{bergayou13} H. Bergeron, J.P. Gazeau, A. Youssef, Phys. Lett. A {\bf 377}  598–605 (2013).

 \bibitem{carnie68}  P. Carruthers and M. M. Nieto,  Phase and angle variables in quantum mechanics,
 \textit{Rev. Mod. Phys.}  \textbf{40} (1968) 411--440

\bibitem{royer96}  A. Royer, Phase states and phase operators for the quantum harmonic oscillator, 
        \textit{Phys. Rev. A} \textbf{53} (1996) 70--108


\bibitem{roberts} J. E. Roberts, The Dirac bra and ket formalism,    {J. Math. Phys.} \textbf{7} (1966)  1097--1104; 
   Rigged Hilbert spaces in quantum mechanics,  \textit{Commun. Math. Phys.} \textbf{3} (1966)  98--119 
   
\bibitem{ant-rhs1} J-P. Antoine, Dirac formalism and symmetry problems in Quantum Mechanics.
      I. General Dirac formalism, {J. Math. Phys.} \textbf{ 10} (1969)  53--69

\bibitem{-jpa_rhs} J-P. Antoine, Quantum mechanics beyond Hilbert space, 
         in \textit{Irreversibility and Causality --- Semigroups and Rigged Hilbert Spaces}, 
        ed. by A. B\"ohm, H-D. Doebner and P. Kielanowski,   Lecture Notes in Physics, vol.  504,
       (Springer-Verlag, Berlin \emph{et al.}, 1998),  pp. 3--33

\bibitem{bohmlect} A. B\"ohm, The  Rigged Hilbert Space in quantum mechanics, in
  \textit{Lectures in  Theoretical Physics\/},   Vol. IX A,  W.A Brittin \emph{et al.} (eds.), pp. 255--315; Gordon \& Breach, New York, 1967
  
\bibitem{reedsimon2} Reed M. and Simon B., 
Methods of Modern Mathematical Physics, II. Fourier Analysis, Self-Adjointness
Volume 2
Academic Press, New York
1975

   \bibitem{stenzel}   M. B. Stenzel, The Segal-Bargmann transform on a symmetric space of compact type, 
{J.  Funct. Analysis} \textbf{165} (1994) 44--58

\bibitem{grosser08} M. Grosser, A note on distribution spaces on manifolds, 
         {Novi Sad Math.  }  \textbf{38} (2008)  121--128

\bibitem{dirac64}  P. A. M. Dirac,  \textit{Lectures on Quantum Mechanics}  (Dover,  New York, 2001)   

\bibitem{bom-klau}   W. R. Bomstad and J. R. Klauder, Linearized quantum gravity using the projection operator formalism, \textit{Class. Quantum Grav.} \textbf{23}  (2006) 5961–-5981

\bibitem{dewitt} B. Dewitt, Quantum theory of gravity. I. The canonical theory,  \textit{Phys. Rev.} \textbf{186}  (1967) 1113--1148

 \bibitem{misner69} C. W. Misner, Quantum Cosmology. I. \textit{Phys. Rev.} \textbf{160}  (1969) 1319--1327

\bibitem{kamin} W. Kami\'nski, J. Lewandowski, and T. Paw{\l}owski, Quantum constraints, Dirac observables and evolution: 
group averaging versus the Schr\"{o}dinger picture in LQC, \textit{Class. Quant. Grav.} \textbf{26} (2009) 245016

\bibitem{rovelli-lect}  C. Rovelli, Zakopane lectures on loop gravity,  in \textit{3rd Quantum Gravity and Quantum Geometry School},
Feb. 28-March 13, 2011, Zakopane, Poland; arXiv:1102.3660v5

\bibitem{wieland12}  W. M. Wieland, Complex Ashtekar variables and reality conditions
for Holst’s action, \textit{Ann. Henri Poincar\'e} \textbf{13} (2012) 425--448

\bibitem{balfrega13} M. Baldiotti, R. Fresneda, and J. P. Gazeau,  About Dirac\&Dirac constraint quantizations, \emph{in progress}

\bibitem{-barracz} A. O. Barut and R. R\c{a}czka, \textit{Theory of Group Representations
       and Applications\/} (PWN, Warszawa, 1977)

\bibitem{aag00} S. T. Ali, J.-P. Antoine, and J.-P. Gazeau,  \textit{Coherent States, Wavelets and their Generalizations\/} (Graduate Texts in Mathematics, 	
Springer, New York,  2000), 2nd Edition to be published, 2014.

\bibitem{alienglis2005}  S.T. Ali  and M. Engli\v{s}     {\it Rev. Math. Phys.} {\bf 17} 391 (2005)

\bibitem{gazeaubook09} Gazeau J.-P. 
Coherent States in Quantum Physics Wiley-VCH, Berlin 2009

\bibitem{-perel2} A. M. Perelomov, \textit{ Generalized Coherent States and their
            Applications\/}  (Springer-Verlag, Berlin, 1986)

\bibitem{-gros_par}  A. Grossmann, Parity operator and quantization of $\delta$-functions, 
         {Commun.~Math. Phys.}  \textbf{48} (1976)  191--194

\bibitem{cahillglauber69} 
K.E. Cahill and R. Glauber, Ordered expansion in Boson Amplitude Operators,   \textit{Phys. Rev.} {\bf 177} 1857-1881 (1969)

\bibitem{magnus66}   Wilhelm Magnus, Fritz Oberhettinger, and Raj~Pal  Soni.
\newblock {\em Formulas and Theorems for
the Special Functions of Mathematical Physics}.
 \newblock Springer-Verlag,  Berlin, Heidelberg and New York, 1966.

\bibitem{cahillglauber69_2} 
K.E. Cahill and R. Glauber, Density Operators and Quasiprobability Distributions,   \textit{Phys. Rev.} {\bf 177} 1882-1902 (1969)

\bibitem{agarwal-wolf70_1}  G. S. Agarwal and E. Wolf, Calculus for Functions of Noncommuting Operators and General Phase-Space Methods in Quantum Mechanics. I. Mapping Theorems and Ordering of Functions of Noncommuting Operators, \textit{Phys. Rev. D} \textbf{2}, 2161-2186 (1970).

\bibitem{agarwal-wolf70_2} G. S. Agarwal and E. Wolf, Calculus for Functions of Noncommuting Operators and General Phase-Space Methods in Quantum Mechanics. II. Quantum Mechanics in Phase Space, \textit{Phys. Rev. D} \textbf{2}, 2187-2205 (1970).
 
 \bibitem{agarwal-wolf70_3} G. S. Agarwal and E. Wolf, Calculus for Functions of Noncommuting Operators and General Phase-Space Methods in Quantum Mechanics. III. A Generalized Wick Theorem and Multitime Mapping, \textit{Phys. Rev. D}\textbf{2}, 2206-2225 (1970).
 	
 
 \bibitem{gold81}   H. Goldstein, C. Poole, and J. Safko, \textit{Classical Mechanics, 3rd ed},
(Addison-Wesley, Reading, MA, 1981)

  \bibitem{galapon02}{E. Galapon, Pauli's theorem and quantum canonical
pairs: The consistency of a bounded, self-adjoint time operator canonically conjugate to a Hamiltonian with
non-empty point spectrum, Proc. R. Soc. Lond. A,   \textbf{458} (2002) 451--472}

 \bibitem{reedsimon} Reed M. and Simon B., 
Methods of Modern Mathematical Physics, I. Functional Analysis
Vol. 1 Academic Press, New York
1972


\bibitem{schwartz61} L. Schwartz, \textit{M\'ethodes math\'ematiques pour les sciences physiques 
(Hermann, Paris,  1961)}

\bibitem{-sudarshan1} E. C. G. Sudarshan,  Equivalence of semiclassical and quantum 
mechanical   descriptions  of statistical light beams,   
{Phys. Rev. Lett.} \textbf{10} (1963) 277--279

\bibitem{boggiatto03} P. Boggiatto and E. Cordero, Anti-Wick quantization of tempered distributions,  
in \textit{Progress in analysis, Vol. I, II, Berlin (2001)}(World Sci. Publ., River Edge, NJ, 2003), pp. 655--662  

\bibitem{boggiatto04} P. Boggiatto, E. Cordero and K. Gr\"ochenig, Generalized Anti-Wick operators with symbols
in distributional Sobolev spaces, Integral Equ. Oper. Theory,  \textbf{48} (2004)  427--442 

\bibitem{hirshfeld02} A. C. Hirshfeld and P. Henselder,  Deformation quantization in the teaching of Quantum Mechanics, arXiv:quant-ph/0208163  (2002)

\bibitem{voros02}  A. Voros,  Wentzel-Kramers-Brillouin method in the Bargmann representation, Phys. Rev. A, 
 \textbf{40} (2002) 6814-6825
 
 \bibitem{daoud03} M. Daoud,  Extended Voros product in the coherent states framework, 
Phys. Lett. A \textbf{309}  (2003) 167--175

\bibitem{alex01}  G. Alexanian, A. Pinzul and A. Stern,  Generalized coherent state approach to star products and applications to the fuzzy sphere, 
Nucl. Phys. B \textbf{600} (2001) 531--547

\bibitem{bal07} A. P. Balachandran, S. Kurkcunoglu and S. Vaidya, \textit{Lectures on Fuzzy and Fuzzy SUSY Physics} (World Scientific, Singapore, 2007)

\bibitem{-glauber1}R. J.~Glauber, 
     The quantum theory of optical coherence,   {Phys. Rev.} \textbf{130}  (1963)  2529--2539; 
      Coherent and incoherent states of radiation field,   ibid. \textbf{131}  (1963) 2766--2788

\bibitem{-klausud}J. R. Klauder and E. C. G. Sudarshan,
   \textit{Fundamentals of Quantum Optics}  (Benjamin, New York, 1968)

\bibitem{cahill65}  K.~E. Cahill,  Coherent-state representations for the photon density, Phys. Rev. \textbf{138}  (1965) B1566--1576 

\bibitem{miller68}  M. M. Miller,  Convergence of the Sudarshan expansion for the diagonal coherent-state weight functional, 
J. Math. Phys.  \textbf{9}  (1968) 1270--1274

\bibitem{kl} J.~R. Klauder, \emph{Fundamentals of Quantum Optics},  (1968).


\bibitem{gelnai} I.M. Gel'fand and M.A. N'aimark, Dokl. Akad. Nauk SSSR\textbf{55} (1947) 570.

\bibitem{aslaklauder} E. W. Aslaksen and J. R. Klauder, Unitary Representations of the Affine Group,  J. Math. Phys. \textbf{15} (1968) 206--211

\bibitem{-grosmor} A. Grossmann and J.~Morlet,    Decomposition of Hardy functions 
      into square integrable wavelets of constant shape,   
         {SIAM J.~Math. Anal.}  \textbf{15} (1984)  723--736.

\bibitem{-gros1} A. Grossmann, J.~Morlet and T. Paul,    Integral transforms associated 
to square   integrable representations. I. General results,   
{J.~Math. Phys.}  \textbf{26} (1985)  2473--2479. 
 
 \bibitem{-gros2} A. Grossmann, J.~Morlet and T. Paul,    Integral transforms 
associated to square  integrable representations. II. Examples,   
{Ann. Inst. H. Poincar\'e} \textbf{45}   (1986)  293--309.    

\bibitem{klauder_Aslaksen70} J. R. Klauder and E. W. Aslaksen, Elementary Model for Quantum Gravity, \textit{Phys. Rev. D} {\bf 2}, 272-276 1970. 

\bibitem{klauder11}  J. R. Klauder, An Affinity for Affine Quantum Gravity,
\textit{Proceedings of the Steklov Institute of Mathematics} {\bf 272}, 169-176 (2011); gr-qc/1003.261

\bibitem{Gesztesy}
F. Gesztesy and W. Kirsch J. Rein. Ang. Math. {\bf 362} 28 (1985).

\bibitem{wcosmo1} H Bergeron, A Dapor, J-P G and P Ma\l kiewicz,
\textit{Wavelet Quantum Cosmology} (2013); arXiv:1305.0653 [gr-qc]

\bibitem{fanuel_Zonetti13}  M. Fanuel and S. Zonetti, Affine Quantization and the Initial Cosmological Singularity, \textit{Eur. Phys. Lett.} {\bf 101},  10001 (2013); gr-qc/1203.4936. 
  


.


.


\end{thebibliography}
\end{document}